\documentclass[]{jfm}

\usepackage{graphicx}
\usepackage{newtxtext}
\usepackage{newtxmath}
\usepackage{natbib}
\usepackage{hyperref}
\hypersetup{
    colorlinks = true,
    urlcolor   = blue,
    citecolor  = black,
}

\newcommand{\RomanNumeralCaps}[1]
\linenumbers
\usepackage{subfig}
 \usepackage[abs]{overpic}
\usepackage{placeins}
\usepackage{tikz}
\usepackage{bm}
\usepackage{booktabs}
\usepackage{placeins}
\usepackage{marginnote}

\newcommand{\dbknote}[1]{\marginnote{\tiny\color{blue}#1}}

\title{Asymptotic analysis of mixing in stratified turbulent flows, and the conditions for an inertial sub-range}

\author{Andrew D. Bragg\aff{1}\corresp{\email{andrew.bragg@duke.edu}}
 \and Stephen M. de Bruyn Kops\aff{2}
 }

\affiliation{\aff{1}Department of Civil and Environmental Engineering, Duke University, Durham, NC 27708, USA
\aff{2}Department of Mechanical and Industrial Engineering, University of Massachusetts Amherst, Amherst, MA 01003, USA
}

\begin{document}

\maketitle

\begin{abstract}
In an important study, Maffioli et al. (J. Fluid Mech., Vol. 794 , 2016) used a scaling analysis to predict that in the weakly stratified flow regime $Fr_h\gg1$ ($Fr_h$ is the horizontal Froude number), the mixing coefficient $\Gamma$ (defined as the ratio of the dissipation rates of potential to kinetic energy) scales as $\Gamma\sim O(Fr_h^{-2})$. Direct numerical simulations confirmed this result, and also indicated that for the strongly stratified regime $Fr_h\ll 1$, $\Gamma\sim O(1)$. Furthermore, the study argued that $\Gamma$ does not depend on the buoyancy Reynolds number $Re_b$, but only on $Fr_h$. We present an asymptotic analysis to predict theoretically how $\Gamma$ should behave for $Fr_h\ll1$ and $Fr_h\gg1$ in the limit $Re_b\to\infty$. To correctly handle the singular limit $Re_b\to\infty$ we perform the asymptotic analysis on the filtered Boussinesq-Navier-Stokes equations, and demonstrate the precise sense in which the inviscid scaling analysis of Billant \& Chomaz (Phys. Fluids, vol. 13 (6), 1645–1651, 2001) applies to viscous flows with $Re_b\to\infty$. The analysis yields $\Gamma\sim O(Fr_h^{-2}(1+Fr_h^{-2}))$ for $Fr_h\gg1$ and $\Gamma\sim O(1+Fr_h^{2})$ for $Fr_h\ll 1$, providing a theoretical basis for the numerical observation made by Maffioli et al, as well as predicting the sub-leading behavior. Our analysis also shows that the Ozmidov scale $L_O$ does not describe the scale below which buoyancy forces are sub-leading, which is instead given by $O(Fr_h^{1/2} L_O)$, and that the condition for there to be an inertial sub-range when $Fr_h\ll 1$ is not $Re_b\gg1$, but the more restrictive condition $Re_b\gg Fr_h^{-4/3}$.
\end{abstract}

\section{Introduction}

In this work we are concerned with the idealized problem of stratified turbulent flows governed by the Boussinesq-Navier-Stokes equations with a spatially and temporally constant background density gradient, and where energy is supplied through an external forcing term in the momentum equation. This idealized problem has been studied extensively as a model for understanding stratified turbulence in environmental flows \citep{waite04,lindborg06a,brethouwer07,waite11,almalkie12a,taylor19}. In such a flow, part of the energy supplied to the turbulent kinetic energy (TKE) field is transferred through reversible stirring processes to the turbulent potential energy (TPE) field. On average, energy in the kinetic and potential fields is transferred to successively smaller scales in the flow until at the smallest scales it is dissipated and mixing takes place.

The fraction of the energy that is on average transferred from the TKE field to the TPE field is governed by purely emergent processes in the turbulent flow, and this makes it challenging to predict. Associated with this is the question of the fraction of energy that is dissipated by the TKE and TPE fields at the small-scales. This is quantified by the so-called mixing coefficient $\Gamma\equiv \langle\chi^*\rangle/\langle\epsilon^*\rangle$ \citep{osborn80}, where $\langle\chi^*\rangle$ is the dissipation rate of TPE, $\langle\epsilon^*\rangle$ is the dissipation rate of TKE, and the superscript $*$ is used throughout to denote a dimensional variable. A challenging question to address is how $\Gamma$ depends on the parameters of the flow \citep{caulfield2020,caulfield2021}, which for the idealized problem under consideration are the horizontal Froude number $Fr_h$, the buoyancy Reynolds number $Re_b$ (concerning which there are two distinct definitions that will be discussed later), and the Prandtl number $Pr$. In this paper we will only consider $Pr=O(1)$, leaving the more general case to future work. One of the reasons for this is that we have recently shown that $Pr$ can lead to surprising and complex effects on stratified turbulence \citep{bragg2023understanding}, and so it is best to leave this additional complication for a subsequent study.

An important study on the parametric dependence of $\Gamma$ in stratified turbulent flows is that of \citet{maffioli16b}. They presented a scaling analysis which predicted that $\Gamma\sim O(Fr_h^{-2})$ in the weakly stratified regime $Fr_h\gg1$, which direct numerical simulations (DNS) confirmed. Their DNS also indicated that in the strongly stratified regime $Fr_h\ll1$, the mixing coefficient asymptotes to $\Gamma\sim O(1)$, although no theoretical analysis was provided to explain this behavior. Moreover, their study argued that $\Gamma$ is independent of $Re_b$ and depends only on $Fr_h$. A set of DNS were presented for which an approximately fixed value $Re_b\gg 1$ was used, while $Fr_h$ was varied over a range spanning weakly to strongly stratified turbulence. The results clearly showed that $\Gamma$ decreased dramatically as $Fr_h$ was increased, however, this only shows that $\Gamma$ is a function of $Fr_h$, and does not demonstrate that it is not also a function of $Re_b$. Moreover, the results in figure 4(b) of \citet{brethouwer07} show that $\Gamma$ decreases strongly with decreasing $Re_b$ when $Re_b< O(10)$. Hence it cannot be true in general that $\Gamma$ is independent of $Re_b$, although it may be for $Re_b\gg1$ (which is possibly the only regime that \citet{maffioli16b} had in mind when arguing that $\Gamma$ is independent of $Re_b$). Furthermore, in the main portion of the results in \citet{maffioli16b}, $Fr_h$ and $Re_b$ are varied simultaneously across the DNS cases, making it difficult to understand to what extent variations in $\Gamma$ across the DNS cases were due to changes in $Fr_h$ only or also due to the changes in $Re_b$. 

The study of \citet{maffioli16b} therefore left open two significant questions. First, to what extent does $\Gamma$ depend on $Fr_h$ as opposed to $Re_b$? Second, how can the result $\Gamma\sim O(1)$ observed in their DNS for $Fr_h\ll1$ be understood on theoretical grounds? To answer the first question, we explore the behavior of $\Gamma$ using an extensive DNS database of stratified turbulence where $Re_b$ is approximately fixed while $Fr_h$ is varied, for a wide range of values of $Re_b$. The study of \citet{Garanaik19} sought to answer the second question and presented a simple scaling analysis that predicts $\Gamma\sim O(1)$ for $Fr_h\ll1$, consistent with the  DNS results of \citet{maffioli16b}. However, the scaling analysis of \citet{Garanaik19} seems problematic. For example, it argues that for $Fr_h\ll1$, $\langle\epsilon^*\rangle \sim O(U_{v,0}^2 N)$, where $U_{v,0}$ is the root-mean-square (r.m.s.) vertical fluid velocity and $N$ is the buoyancy frequency. Using the scaling results of \citet{billant01,brethouwer07} we have $U_{v,0}\sim O(Fr_h U_{h,0})$, with $Fr_h\equiv U_{h,0}/(L_{h,0}N)$, where $U_{h,0}$ is the r.m.s. horizontal fluid velocity and $L_{h,0}$ is the horizontal integral length of the horizontal velocity field. Using these in $\langle\epsilon^*\rangle \sim O(U_{v,0}^2 N)$ yields $\langle\epsilon^*\rangle \sim O(Fr_h U_{h,0}^3/L_{h,0})$ which is fundamentally inconsistent with the classical result $\langle\epsilon^*\rangle \sim O(U_{h,0}^3/L_{h,0})$ which is regarded as a well established result for strongly stratified turbulent flows \citep{riley12,maffioli16}. The reason the analysis of \citet{Garanaik19} nevertheless correctly predicts $\Gamma\sim O(1)$ for $Fr_h\ll1$ is because their scaling correctly predicts that $\langle\epsilon^*\rangle$ and $\langle\chi^*\rangle$ are of the same order in this regime, despite the fact that the scaling estimates for $\langle\epsilon^*\rangle$ and $\langle\chi^*\rangle$ themselves are incorrect. 

In view of these issues with the proposed theoretical explanation for the result $\Gamma\sim O(1)$ for $Fr_h\ll1$ given by \citet{Garanaik19}, we develop a new asymptotic analysis of $\Gamma$ that predicts its dependence on $Fr_h$ in the limit $Re_b\to \infty$. One of the regimes of interest is the strongly stratified turbulence regime where $Fr_h\ll1$ and $Re_b\gg 1$. The seminal study of \citet{billant01} explored the dynamics of stratified flows in the regime $Fr_h\ll1$ for inviscid fluids and discovered a new scaling regime that arises due to an emergent self-similarity of the flow in this regime. \citet{brethouwer07} extended the analysis to the case of viscous fluids and argued that when $Re_b\gg 1$, the behavior for $Fr_h\ll1$ reduces to the self-similar scaling regime identified by \citet{billant01}. However, this conclusion is problematic because the limit $Re_b\to\infty$ is singular. We instead perform an asymptoptic analysis on the filtered Boussinesq-Navier-Stokes equations in the strongly stratified turbulent regime, which allows the singular limit $Re_b\to\infty$ to be handled correctly. Our analysis then reveals the precise sense in which the inviscid scaling analysis of \citet{billant01} applies to flows where viscous effects are important for at least a sub-set of flow scales. This analysis then enables us to construct asymptotic predictions for $\Gamma$ in the limit $Re_b\to\infty$ for both the $Fr_h\ll1$ and $Fr_h\gg 1$ regimes.

The outline of the paper is as follows. In \S\ref{SSlimit} we explain how an analysis of the filtered governing equations can be used to correctly handle the singular infinite Reynolds number limit for the simpler case of isotropic turbulence. In \S\ref{AsymPred} we apply this approach and develop a new asymptotic analysis of the filtered Boussinesq-Navier-Stokes equations, leading to predictions for the asymptotic behavior of the mixing coefficient $\Gamma$. In \S\ref{Loz} we consider the definition of the Ozmidov scale and the conditions required for an inertial sub-range in strongly stratified turbulent flows. In \S\ref{DNS} we summarize the extensive DNS database that is used to test the theoretical predictions, and in \S\ref{RandD} we present and discuss the results. Finally, in \S\ref{Conc} we provide conclusions and identify important steps for future work.

\section{Scaling in the singular high Reynolds number limit}\label{SSlimit}

We are interested in the high Reynolds number limit, and this can lead to complications in a scaling analysis because of the singular nature of this limit. Since the reader may not be familiar with the issues, we discuss them in the simpler context of neutrally buoyant, incompressible, isotropic turbulence, and we show how a filtering approach can enable the singular limit to be handled correctly. We will then extend these ideas in the next section to consider the more complicated case of stably stratified turbulent flows, enabling asymptotic predictions for $\Gamma$ to be derived.

The Navier-Stokes equation is
\begin{align}
\partial_t^*{\bm{u}}^*+({\bm{u}}^*\bm{\cdot}\bm{\nabla}^*){\bm{u}}^*=&-(1/\rho)\bm{\nabla}^*{p}^*+\nu\nabla^{*^2}{\bm{u}}^*,\label{NSE}
\end{align}
where $\bm{u}$ is the fluid velocity, $p$ the fluid pressure, $\rho$ the fluid density, $\nu$ the fluid kinematic viscosity, and the superscript $*$ denotes that the flow variable is dimensional. Scaling variables using the integral length scale $L_0$ and root-mean-square (r.m.s.) velocity $U_0$, and pressure using $\rho U_0^2$, we obtain the dimensionless form of the equation
\begin{align}
\partial_t{\bm{u}}+({\bm{u}}\bm{\cdot}\bm{\nabla}){\bm{u}}=&-\bm{\nabla}{p}+\frac{1}{Re}\nabla^2{\bm{u}},\label{Euler}
\end{align}
where $Re\equiv U_0L_0/\nu$ is the Reynolds number. Taking the limit $Re\to \infty$ would seem to suggest that in this limit viscous forces become irrelevant and the equation reduces to the inviscid Euler equation. If this were true then (assuming the absence of singularities in the solutions) it would suggest that turbulent flows with $Re\to \infty$ conserve kinetic energy. However, this conclusion is fundamentally inconsistent with standard turbulence theory. The issue arises because the Navier-Stokes equation is singular in the limit $Re\to \infty$, and associated with this is the fact that the viscous term cannot be appropriately scaled using $U_0$ and $L_0$. 

Let us instead assume that the variables in \eqref{NSE} scale with $\mathcal{U}_{\mathcal{L}}$ and $\mathcal{L}$. In a turbulent flow with $Re\to\infty$ there are a wide range of scales in the flow and in principle $\mathcal{U}_{\mathcal{L}}\in[u_\eta, U_0]$, $\mathcal{L}\in[\eta, L_0]$, where $u_\eta$ and $\eta$ are the Kolmogorov velocity and length scales. Using $\mathcal{U}_{\mathcal{L}}$ and $\mathcal{L}$ the scaled Navier-Stokes equation becomes
\begin{align}
\partial_t{\bm{u}}+({\bm{u}}\bm{\cdot}\bm{\nabla}){\bm{u}}=&-\bm{\nabla}{p}+\frac{1}{\mathcal{R}e}\nabla^2{\bm{u}},
\end{align}
where $\mathcal{R}e\equiv \mathcal{L}\mathcal{U}_{\mathcal{L}}/\nu$. In the absence of additional information, among their ranges of possible values it is not clear what the particular values of $\mathcal{U}_{\mathcal{L}}$ and $\mathcal{L}$ should be in order for $\|\nabla^2{\bm{u}}\|\sim O(1)$ to hold. As a result, taking the limit $\mathcal{R}e\to\infty$ is problematic. Indeed, if $\mathcal{U}_{\mathcal{L}}\sim O(u_\eta)$ and $\mathcal{L}\sim O(\eta)$ are the appropriate choices then due to the definitions of $u_\eta$ and $\eta$, $\mathcal{R}e\sim O(1)$ and so taking the limit $\mathcal{R}e\to\infty$ would not be valid. 

Despite these points, we do nevertheless expect, based on standard turbulence theory, that at the large-scales of the flow the direct influence of viscous forces will be negligible. It could be argued that the fact that the viscous term vanishes for $Re\to\infty$ when the flow variables are scaled using $U_0$ and $L_0$ only indicates that the \emph{large scales} of a turbulent flow obey the Euler equation. However, as discussed below, this inference is not correct. 

A more precise way to address these issues is to consider the filtered Navier-Stokes equation \citep{leonard74,germano92,eyink05}, with variables scaled using $\mathcal{U}_{\mathcal{L}}$ and $\mathcal{L}$
\begin{align}
\partial_t\widetilde{\bm{u}}+(\widetilde{\bm{u}}\bm{\cdot}\bm{\nabla})\widetilde{\bm{u}}=&-\bm{\nabla}\widetilde{p}+\frac{1}{\mathcal{R}e}\nabla^2\widetilde{\bm{u}}-\frac{\mathcal{L}}{\mathcal{U}_{\mathcal{L}}^2}\bm{\nabla}^*\bm{\cdot}\bm{\tau}^*,\label{CGEuler}
\end{align}
where $\widetilde{(\cdot)}$ denotes a filtered variable, and $\bm{\tau}\equiv \widetilde{\bm{u}\bm{u}}-\widetilde{\bm{u}}\widetilde{\bm{u}}$ is the sub-grid stress tensor. Since the velocities in this equation are filtered, then $\mathcal{U}_{\mathcal{L}}$ and $\mathcal{L}$ fall into the restricted ranges $\mathcal{U}_{\mathcal{L}}\geq \mathcal{U}_{\Delta}$, $\mathcal{L}\geq\Delta$, where $\mathcal{U}_{\Delta}$ is the smallest velocity scale present in the filtered velocity field, and $\Delta$ is the filter length. 

Once again, in the absence of additional information, among their ranges of possible values it is not clear what the particular values of $\mathcal{U}_{\mathcal{L}}$ and $\mathcal{L}$ should be in order to generate the correct scaling. However, if we consider $\Delta/\eta\gg1$ then $ \mathcal{U}_{\Delta}\gg u_\eta$ and hence the minimum value that $\mathcal{R}e$ could take is $\mathcal{R}e=\Delta  \mathcal{U}_{\Delta}/\nu\gg 1$. In this case, taking the limit $\mathcal{R}e\to\infty$ is well defined and corresponds to $\Delta/\eta\to\infty$, for which the viscous term in the filtered equation can be ignored, and all the direct effects of the viscous force will be isolated to the sub-grid flow. Note however that in this case the filtered Navier-Stokes equation does not reduce to the Euler equation, but to the filtered Euler equation due to the sub-grid stress term, such that the large-scales of a high Reynolds number turbulent flow do not obey the Euler equation, but the filtered Euler equation. According to this, the filtered flow loses energy, not due to viscous stress, but due to the sub-grid stress which causes energy to pass to the sub-grid flow on average (in a three dimensional flow), i.e. due to the energy cascade.

\citet{brethouwer07} scale the Boussinesq-Navier-Stokes equations using large-scale horizontal length $L_{h,0}$ and r.m.s. horizontal velocity $U_{h,0}$ and conclude that for $Fr_h\equiv U_{h,0}/(L_{h,0} N)\ll 1$ and $Re_b\equiv Fr_h^2 Re_h\gg 1$ (where $Re_h\equiv L_{h,0} U_{h,0}/\nu$) the viscous and diffusive terms in the equations can be neglected, and the equations become equivalent to the inviscid equations analyzed in \citet{billant01}. This conclusion is problematic in the same way that it was shown to be problematic to assume that the term $(1/Re)\nabla^2{\bm{u}}$ in \eqref{Euler} is negligible for $Re\to\infty$. To carefully handle the high Reynolds number limit in the context of stratified flows we could instead consider the filtered Boussinesq-Navier-Stokes equations, and this is what we will do in the next section.

In order to perform a scaling analysis on \eqref{CGEuler}, we must make particular choices for the velocity and length scales. To scale the filtered velocities we can simply use the root-mean-square value $\|\widetilde{\bm{u}^*}\|\sim O(U)$ where $U^2\equiv \langle\|\widetilde{\bm{u}^*}\|^2 \rangle$, and we note that by definition $U^2= U_0^2\equiv \langle\|{\bm{u}^*}\|^2 \rangle$ for $\Delta=0$. However, in general, the contribution to $\|\widetilde{\bm{u}^*}\|$ and $\|\bm{\nabla}^*\widetilde{\bm{u}^*}\|$ from a given scale in the filtered flow will be different, e.g. larger scales of the filtered flow may contribute most to $\|\widetilde{\bm{u}^*}\|$, while smaller scales of the filtered flow may contribute most to $\|\bm{\nabla}^*\widetilde{\bm{u}^*}\|$. Therefore, for terms involving derivatives of the filtered velocity (which includes the pressure gradient term due to the pressure Poisson equation), the velocity scale will be instead chosen to be $\mathcal{U}$ such that $\|\bm{\nabla}^*\widetilde{\bm{u}^*}\|\sim O(\mathcal{U}/\ell)$, $\|\partial_{t^*}\widetilde{\bm{u}^*}\|\sim O(\mathcal{U}^2/\ell)$, and $\|\bm{\nabla}^*\widetilde{p}^*\|\sim O(\mathcal{U}^2/\ell)$, where $\ell$ is a length that scales the derivatives and $\ell\sim O(\Delta)$. The relationship between $\mathcal{U}$ and $U$ will emerge from the analysis of the equations itself. The sub-grid stress will also be assumed to scale in the same way as the filtered inertial terms, i.e. $\|\bm{\nabla}^*\bm{\cdot}\bm{\tau}^*\|\sim O(\mathcal{U}^2/\ell)$. Finally, to consider a statistically stationary regime we introduce an isotropic forcing term into the equations, whose filtered contribution is $\widetilde{\bm{F}^*}$, and we assume the scaling $\|\widetilde{\bm{F}^*}\| \sim O(U^2/L)$, where $L$ is the integral lengthscale of the filtered velocity field. This choice is suitable since it is the forcing that is driving the filtered velocity field in the first place, and this forcing is assumed to be confined to the largest scales of the flow.

With these choices, \eqref{CGEuler} becomes
\begin{align}
\partial_t\widetilde{\bm{u}}+(\widetilde{\bm{u}}\bm{\cdot}\bm{\nabla})\widetilde{\bm{u}}=&-\bm{\nabla}\widetilde{p}+\frac{1}{\mathcal{R}e}\nabla^2\widetilde{\bm{u}}-\bm{\nabla}\bm{\cdot}\bm{\tau}+\frac{\mathcal{\ell}}{\mathcal{U}^2}\frac{U^2}{L}\widetilde{\bm{F}},\label{CGEuler2}
\end{align}
where now $\mathcal{R}e= \ell\mathcal{U}/\nu$, and in the following we will consider scales $\ell$ for which the limit $\mathcal{R}e\to \infty$ can be taken (i.e. the inertial range).

From \eqref{CGEuler2} the large-scale and small-scale TKE equations for a statistically stationary and homogeneous flow can be derived, and when multiplied by $\ell/\mathcal{U}^3$ they become (for $\mathcal{R}e\to \infty$)
\begin{align}
0&=-\langle \Pi_K\rangle+\frac{\ell}{\mathcal{U}^3}\frac{U^3}{L}\langle\widetilde{\bm{F}}\bm{\cdot}\widetilde{\bm{u}}\rangle,\\
0&=-\langle \Pi_K\rangle+\frac{\ell}{\mathcal{U}^3}2\nu\langle\widetilde{\|\bm{S}^*\|^2}-\|\widetilde{\bm{S}^*}\|^2\rangle -\frac{\ell}{\mathcal{U}^3}\langle\widetilde{\bm{F}^*\bm{\cdot}\bm{u}^*}-\widetilde{\bm{F}^*}\bm{\cdot}\widetilde{\bm{u}^*} \rangle,
\end{align}
where $\Pi_K^*\equiv -\bm{\tau}^*\bm{:}\bm{\nabla}^*\widetilde{\bm{u}}^*$ is the inter-scale TKE flux, and $\widetilde{\bm{S}}^*\equiv (\bm{\nabla}^*\widetilde{\bm{u}}^*+[\bm{\nabla}^*\widetilde{\bm{u}}^*]^\top)/2$ is the filtered strain-rate tensor. 

For a homogeneous turbulent flow $2\nu\langle\widetilde{\|\bm{S}^*\|^2}\rangle=\langle\epsilon^*\rangle$, and therefore the small-scale TKE dissipation rate can be re-written as $2\nu\langle\widetilde{\|\bm{S}^*\|^2}-\|\widetilde{\bm{S}^*}\|^2\rangle=\langle\epsilon^*\rangle-2\nu \langle\|\widetilde{\bm{S}^*}\|^2\rangle$. In the limit $\mathcal{R}e\to \infty$, $(\ell/\mathcal{U}^3)2\nu \langle\|\widetilde{\bm{S}^*}\|^2\rangle=0$.

Assuming that $\ell\ll L_0$ and that the forcing only acts at scales $O(L_0)$ (as is usually the case in DNS) then
\begin{align}
\frac{\ell}{\mathcal{U}^3}\langle\widetilde{\bm{F}^*\bm{\cdot}\bm{u}^*}-\widetilde{\bm{F}^*}\bm{\cdot}\widetilde{\bm{u}^*} \rangle\approx 0.
\end{align}
The large and small-scale TKE equations then become
\begin{align}
0&=-\langle \Pi_K\rangle+\frac{\ell}{\mathcal{U}^3}\frac{U^3}{L}\langle\widetilde{\bm{F}}\bm{\cdot}\widetilde{\bm{u}}\rangle,\\
0&=-\langle \Pi_K\rangle+\frac{\ell}{\mathcal{U}^3}\langle\epsilon^*\rangle,
\end{align}
and from these we obtain the scaling relationships
\begin{align}
\mathcal{U}&\sim O\Big(U(\ell/L)^{1/3}\Big),\label{K41_Ul}\\
\langle\epsilon^*\rangle&\sim O\Big(U^3/L\Big).
\end{align}
The former result does not explicitly determine the $\ell$-dependence of $\mathcal{U}$ because the $\ell$-dependence of $U$ and $L$ are not yet known. To determine these in terms of $U_0$ and $L_0$ (which are independent of $\ell$), we note that for a statistically homogeneous flow
\begin{align}
 \langle\|{\bm{u}^*}\|^2 \rangle= \langle\|\widetilde{\bm{u}^*}\|^2 \rangle+\langle  \widetilde{\|\bm{u}^*\|^2} - \|\widetilde{\bm{u}^*}\|^2 \rangle,
\end{align}
and that $\langle  \widetilde{\|\bm{u}^*\|^2} - \|\widetilde{\bm{u}^*}\|^2 \rangle\geq 0$ for non-negative filter kernels \citep{vreman_geurts_kuerten_1994}, implying $\langle\|{\bm{u}^*}\|^2 \rangle
\geq\langle\|\widetilde{\bm{u}^*}\|^2 \rangle$ (i.e. $U_0^2\geq U^2$). Since $\mathcal{U}$ corresponds to a velocity scale in the filtered field, and since the velocity scales in the small-scale field are less than or equal to those in the filtered field, then the small-scale contribution must satisfy $\langle  \widetilde{\|\bm{u}^*\|^2} - \|\widetilde{\bm{u}^*}\|^2 \rangle\leq O(\mathcal{U}^2)$. Using \eqref{K41_Ul} we therefore obtain 
\begin{align}
 \langle\|{\bm{u}^*}\|^2 \rangle\leq O\Big(U^2\Big(1+(\ell/L)^{2/3}\Big)\Big),
\end{align}
and with the definition $U_0^2\equiv \langle\|{\bm{u}^*}\|^2 \rangle$ together with $U_0^2\geq U^2$ this leads to
\begin{align}
U_0^2\geq U^2\geq O\Big(U_0^2\Big(1+(\ell/L)^{2/3}\Big)^{-1}\Big).
\end{align}
This shows that for $\ell\ll L$, $U\sim O(U_0)$. The fact that $U$ converges to $U_0$ as $\ell$ is reduced reflects the familiar idea that in a high Reynolds number isotropic turbulent flow, it is the large-scales that make the dominant contribution to the total TKE $U_0^2/2$. Similar reasoning can also be used to show that $L\sim O(L_0)$ when $\ell\ll L$, reflecting the fact that in a high Reynolds number turbulent flow, the integral lengthscale is dominated by the large scales in the flow. As a consequence, the results obtained previously become 
\begin{align}
\mathcal{U}&\sim O\Big(U_0(\ell/L_0)^{1/3}\Big),\\
\langle\epsilon^*\rangle&\sim O(U_0^3/L_0),
\end{align}
which correspond to Kolmogorov scaling for velocities in the inertial range, and Taylor scaling for the TKE dissipation rate \citep{kolmogorov41,taylor35,pope00,davidsonb}, respectively. 

The results obtained above were derived for the limit $\mathcal{R}e\to \infty$. However, if we extrapolate (in the spirit of the method of matched asymptotics) the results down to the scale at which $\mathcal{R}e\sim O(1)$ we find that this scale is given by $\ell\sim O((\nu^3/\langle\epsilon^*\rangle)^{1/4})$ with corresponding velocity scale $\mathcal{U}\sim O((\nu\langle\epsilon^*\rangle)^{1/4})$, and these are nothing other than the Kolmogorov length $\eta$ and velocity $u_\eta$ scales. This demonstrates that the choice of velocity and length scales chosen earlier to scale the filtered equation leads to the well established Kolmogorov results for the mean-field behavior of turbulent flows in both the inertial and dissipation ranges.

Extensions of the scaling and method just presented will be used in what follows when analyzing the filtered Boussinesq-Navier-Stokes equations, except that the scaling in the vertical and horizontal directions will be distinguished since stably stratified flows are anisotropic.

\section{Asymptotic analysis of mixing in stratified flows}\label{AsymPred}

We consider flows governed by the forced Boussinesq-Navier-Stokes equations, and given the anisotropy of the flow due to stratification, we write separate equations for the horizontal velocity ${\bm{u}_h}^*$ and vertical velocity ${{u}_z}^*$. With the variable fluid density decomposed as $\rho=\rho_r+z\nabla_z\langle\rho\rangle+\varrho$, where $\rho_r$ is a reference density, $\nabla_z\langle\rho\rangle<0$ is the constant mean density gradient, and $\varrho$ is the fluctuation about the mean density $\langle\rho\rangle=\rho_r+z\zeta$, the equations are written as
\begin{align}
\bm{\nabla}_h^*\bm{\cdot}{\bm{u}_h}^*&=-\nabla_z^*{{u}_z}^*,
\\
\partial_t^*{\bm{u}_h}^*+({\bm{u}_h}^*\bm{\cdot}\bm{\nabla}_h^*){\bm{u}_h}^*+({{u}_z}^*\nabla_z^*){\bm{u}_h}^*&=-(1/\rho_r)\bm{\nabla}_h^*{p}^*+\nu\nabla_h ^{*^2}{\bm{u}_h}^*+\nu\nabla_z ^{*^2}{\bm{u}_h}^*+{\bm{F}}_h^*,
\\
\partial_t^*{{u}_z}^*+({\bm{u}_h}^*\bm{\cdot}\bm{\nabla}_h^*){{u}_z}^*+({{u}_z}^*\nabla_z^*){{u}_z}^*&=-(1/\rho_r)\nabla_z^*{p}^*+\nu\nabla_h ^{*^2}{{u}_z}^*+\nu\nabla_z ^{*^2}{{u}_z}^*- N\phi^*,
\\
\partial_t^*{\phi}^*+({\bm{u}_h}^*\bm{\cdot}\bm{\nabla}_h^*)\phi^*+({{u}_z}^*\nabla_z^*)\phi^*&=\kappa\nabla_h ^{*^2}{\phi}^*+\kappa\nabla_z ^{*^2}{\phi}^* +N {u}_z^*,
\end{align}
where $\phi^*\equiv g\varrho^*/(N/\rho_r)$ is a variable that is proportional to the fluctuating density $\varrho^*$ and has dimensions of a velocity, $g$ is the gravitational acceleration, $N\equiv \sqrt{-g\nabla_z\langle\rho\rangle/\rho_r}$ is the buoyancy frequency, and $\kappa$ is the thermal diffusivity. In order to ensure that the vertical dynamics are purely emergent, only the horizontal momentum equation is forced as in \citet{lindborg06a,brethouwer07}, with the forcing chosen to generate a flow that is statistically axisymmetric about $\bm{e}_z$ (the unit vector in the vertical direction).

Following \citet{zhao2023measuring}, we will use an anisotropic filtering operator in order to distinguish between the horizontal and vertical motions of the flow which is important for stratified flows. In particular, for an arbitrary field variable $\bm{a}(\bm{x}_h,\bm{z},t)$, where $\bm{x}_h$ is the position vector in the horizontal plane, $\bm{z}\equiv z\bm{e}_z$ with $z$ the vertical coordinate, we define the filtering operation as 
\begin{align}
\widetilde{\bm{a}}(\bm{x}_h,\bm{z},,t)\equiv \int\int \bm{a}(\bm{x}_h+\bm{x}^\prime_h,\bm{z}+\bm{z}^\prime,t)\mathcal{G}_{\Delta_h}(\|\bm{x}^\prime_h\|) \mathcal{G}_{\Delta_v}(\|\bm{z}^\prime\|)\, d\bm{x}^\prime_h\, d\bm{z}^\prime,    
\end{align}
where the horizontal $\mathcal{G}_{\Delta_h}$ and vertical $\mathcal{G}_{\Delta_v}$ filtering kernels satisfy the normalization property $\int\mathcal{G}_{\Delta_h}(\|\bm{x}^\prime_h\|)\, d\bm{x}^\prime_h=1$ and $\int\mathcal{G}_{\Delta_v}(\|\bm{z}^\prime\|)\, d\bm{z}^\prime=1$, and $\Delta_h$ and $\Delta_v$ denote the filtering lengths in the horizontal and vertical directions, respectively. The filtering kernels must be strictly non-negative in order to preserve the non-negativity of the TKE and TPE in the equations that follow \citep{vreman_geurts_kuerten_1994}. By varying $\Delta_h,\Delta_v$ we can consider the dynamics of stratified turbulence at different scales in the flow. 

We consider the limit $Re_b\to \infty$, and the scaling we will use for the terms in the filtered Boussinesq-Navier-Stokes equations are extensions of those discussed in the previous section for isotropic turbulence.  We scale the horizontal filtered velocity with its r.m.s value $U_{h}\equiv \sqrt{\langle\|\widetilde{\bm{u}_h}^*\|^2 \rangle}$, the vertical filtered velocity with its r.m.s value $U_{v}\equiv \sqrt{\langle |\widetilde{{u}_z}^*|^2 \rangle}$, and the filtered density variable with its r.m.s value $Q\equiv \sqrt{\langle |\widetilde{\phi}^*|^2 \rangle}$. For terms in the equations involving derivatives of filtered variables, the horizontal and vertical velocity scales in the derivatives are $\mathcal{U}_{h}$ and $\mathcal{U}_{v}$, and for terms involving gradients of the density, the density scale is $\mathcal{Q}$. Horizontal and vertical derivative operators will be taken to scale with the inverse of the lengths $\ell_h\sim O(\Delta_h)$ and $\ell_v\sim O(\Delta_v)$, respectively, and the time derivative with $\mathcal{U}_{h}/\ell_h$. Due to the decomposition between horizontal and vertical directions, the pressure is scaled using $\rho_r(\mathcal{U}_{h}^2+\mathcal{U}_{v}^2)$. For later use, we note that when $\ell_h\sim O(L_h)$ and $\ell_v\sim O(L_v)$ the flow is effectively single-scale (in the sense that the scale separation is small, although finite) and therefore in this case $\mathcal{U}_{h}\sim O(U_h)$, $\mathcal{U}_{v}\sim O(U_v)$, and $\mathcal{Q}\sim O(Q)$. Finally, the horizontal forcing ${\bm{F}}_h^*$ will be assumed to act only at the largest scales of the flow and scale with $U_{h}^2/L_{h}$, where $L_{h}$ is the horizontal integral length of the horizontal filtered velocity field.

Using these, the scaled, filtered Boussinesq-Navier-Stokes equations are (assuming that the Prandtl number is $O(1)$)
\begin{align}
&\bm{\nabla}_h\bm{\cdot}\widetilde{\bm{u}_h}=-\zeta\nabla_z\widetilde{{u}_z},\label{cont_eq}\\
\begin{split}
&\partial_t\widetilde{\bm{u}_h}+(\widetilde{\bm{u}_h}\bm{\cdot}\bm{\nabla}_h)\widetilde{\bm{u}_h}+\zeta(\widetilde{{u}_z}\nabla_z)\widetilde{\bm{u}_h}\\
&=-(1+\zeta^2\delta^2)\bm{\nabla}_h\widetilde{p}+\frac{1}{\mathcal{R}_h}\Big[\nabla_h^2\widetilde{\bm{u}_h}+\frac{1}{\delta^2}\nabla_z^2\widetilde{\bm{u}_h}\Big]+\frac{U_{h}^2}{L_{h}}\frac{\ell_h}{\mathcal{U}_{h}^2}\widetilde{\bm{F}}_h-\bm{\nabla}_h\bm{\cdot}\bm{\tau}_{hh}-\zeta{\nabla}_z\bm{\tau}_{zh},\label{uh_eq}
\end{split}
\\
\begin{split}
&\partial_t\widetilde{{u}_z}+(\widetilde{\bm{u}_h}\bm{\cdot}\bm{\nabla}_h)\widetilde{{u}_z}+\zeta(\widetilde{{u}_z}\nabla_z)\widetilde{{u}_z}\\
&=-\Big(\frac{1}{\zeta\delta^2}+\zeta\Big){\nabla}_z\widetilde{p}-\alpha\widetilde{\phi}+\frac{1}{\mathcal{R}_h}\Big[\nabla_h^2\widetilde{{u}_z}+\frac{1}{\delta^2}\nabla_z^2\widetilde{{u}_z}\Big]-\bm{\nabla}_h\bm{\cdot}\bm{\tau}_{zh}-\zeta{\nabla}_z\tau_{zz},\label{uz_eq}
\end{split}
\\
\begin{split}
&\partial_t\widetilde{\phi}+(\widetilde{\bm{u}_h}\bm{\cdot}\bm{\nabla}_h)\widetilde{\phi}+\zeta(\widetilde{{u}_z}\nabla_z)\widetilde{\phi}\\
&= \frac{U_{v}}{\mathcal{F}_h\mathcal{Q}}\widetilde{u}_z+\frac{1}{\mathcal{R}_h}\Big[\nabla_h^2\widetilde{\phi} +\frac{1}{\delta^2}\nabla_z^2\widetilde{\phi}\Big]-\bm{\nabla}_h\bm{\cdot}\bm{\Sigma}_h-\zeta\nabla_z\Sigma_z,\label{phi_eq}
\end{split}
\end{align}
where $\delta\equiv\ell_v/\ell_h$, $\zeta\equiv\mathcal{U}_{v}/(\delta \mathcal{U}_{h})$, $\alpha\equiv Q/(\mathcal{F}_h\mathcal{U}_{v})$, $\mathcal{R}_h\equiv \ell_h \mathcal{U}_{h}/\nu$ is the scale-dependent horizontal Reynolds number, and $\mathcal{F}_h\equiv \mathcal{U}_{h}/(\ell_h N)$ is the scale-dependent horizontal Froude number. 

The sub-grid stress terms are $\bm{\tau}_{hh}^*\equiv\widetilde{\bm{u}_h^*\bm{u}_h^*}-\widetilde{\bm{u}_h}^*\widetilde{\bm{u}_h}^*$, $\bm{\tau}_{zh}^*\equiv \widetilde{{u}_z^*\bm{u}_h^*}-\widetilde{{u}_z}^*\widetilde{\bm{u}_h}^*$, $\tau_{zz}^*\equiv \widetilde{{u}_z^*{u}_z^*}-\widetilde{{u}_z}^*\widetilde{{u}_z}^*$, $\bm{\Sigma}_h^*\equiv \widetilde{\bm{u}_h^*\phi^*}-\widetilde{\bm{u}_h}^*\widetilde{\phi}^*$, $\Sigma_z^*\equiv \widetilde{{u}_z^*\phi^*}-\widetilde{{u}_z}^*\widetilde{\phi}^*$, and in the equations above these have been assumed to scale as
\begin{align}
\bm{\nabla}_h^*\bm{\cdot}\bm{\tau}_{hh}^* +{\nabla}_z^*\bm{\tau}_{zh}^* &\sim O\Big(\frac{\mathcal{U}_{h}^2}{\ell_h}+\frac{\mathcal{U}_{v}\mathcal{U}_{h}}{\ell_v}\Big),\label{SGS_scaling1}\\
\bm{\nabla}_h^*\bm{\cdot}\bm{\tau}_{zh}^* +{\nabla}_z^*\tau_{zz}^* &\sim O\Big(\frac{\mathcal{U}_{v}\mathcal{U}_{h}}{\ell_h}+\frac{\mathcal{U}_{v}^2}{\ell_v}\Big),\label{SGS_scaling2}\\
\bm{\nabla}_h^*\bm{\cdot}\bm{\Sigma}_h^*+{\nabla}_z^*\Sigma_z^* &\sim O\Big(\frac{\mathcal{U}_{h} \mathcal{Q}}{\ell_h}+\frac{\mathcal{U}_{v}\mathcal{Q}}{\ell_v}\Big).\label{SGS_scaling3}
\end{align}
%
When $\min[\mathcal{R}_h,\mathcal{R}_h\delta^2]\to\infty$ the viscous and diffusive terms in the filtered equations vanish, showing that the large-scales of a stratified flow with $\min[\mathcal{R}_h,\mathcal{R}_h\delta^2]\to\infty$ do not obey to leading order the Boussinesq-Euler equations that were studied by \citet{billant01}, but rather they obey the {filtered} Boussinesq-Euler equations.

\subsection{Weakly stratified regime}

The buoyancy term in \eqref{uz_eq} is $O(\alpha)$, and the weakly stratified regime corresponds to $\alpha\ll1\forall \ell_h,\ell_v$. Since the equations are regular in the limit $\alpha\to 0$, this suggest that for $\alpha\ll 1$ we may use the regular perturbation expansion $\widetilde{{u}_z}=\widetilde{{u}_z}^{[0]}+\alpha\widetilde{{u}_z}^{[1]}+O(\alpha^{2})$, where the superscript ${[0]}$ on a variable denotes that the variable corresponds to the solution for $\alpha\to 0$, and we use corresponding expansions for the other variables. We will also assume in the analysis that $\mathcal{R}_h\to\infty$ (it will be seen that for the weakly stratified regime $\delta\sim O(1)$ to leading order, and therefore $\mathcal{R}_h\to\infty$ also implies the limit $\mathcal{R}_h\delta^2\to\infty$).

Equations for the average large-scale horizontal TKE $E_{K,h}\equiv \langle\|\widetilde{\bm{u}_h}\|^2\rangle/2$, large-scale vertical TKE $E_{K,v}\equiv\langle|\widetilde{{u}_z}|^2\rangle/2$ and large-scale TPE $E_P\equiv \langle\widetilde{\phi}^2\rangle/2$ can be derived from \eqref{uh_eq}, \eqref{uz_eq}, and \eqref{phi_eq} (see \citet{zhang22}). Re-arranging these to give equations for the energy transfer terms, inserting into the right hand sides of these equations the perturbation expansions, then for a statistically stationary, homogeneous flow we obtain (for $\mathcal{R}_h\to\infty$)
\begin{align}
\begin{split}
\langle\Pi_{K,hh}\rangle+\zeta\langle\Pi_{K,zh}\rangle=&(1+\zeta^2\delta^2)\langle \widetilde{p}^{[0]}\bm{\nabla}_h\bm{\cdot}\widetilde{\bm{u}_h}^{[0]}\rangle +\gamma\langle \widetilde{\bm{u}_h}^{[0]}\bm{\cdot}\widetilde{\bm{F}}_h^{[0]}\rangle+c_1\alpha,
\end{split}
\\
\begin{split}
\langle\Pi_{K,hz}\rangle+\zeta\langle\Pi_{K,zz}\rangle=&\Big(\frac{1}{\zeta\delta^2}+\zeta\Big)\langle \widetilde{p}^{[0]}\nabla_z\widetilde{{u}_z}^{[0]}\rangle +c_2\alpha,
\end{split}
\\
\begin{split}
\langle\Pi_{P,hz}\rangle+\zeta\langle\Pi_{P,zz}\rangle=&\beta \langle\widetilde{\phi}^{[0]}\widetilde{{u}_z}^{[0]}\rangle +c_3\alpha,
\end{split}
\end{align}
where $\beta\equiv {U_{v}Q/(\mathcal{F}_h\mathcal{Q}^2})=\alpha U_v\mathcal{U}_v\mathcal{Q}^{-2}$, $\gamma\equiv \ell_h U_{h}^3/(L_{h}\mathcal{U}_{h}^3)$, $\Pi_{K,hh}\equiv -\bm{\tau}_{hh}\bm{:}\bm{\nabla}_h\widetilde{\bm{u}_h}$, $\Pi_{K,zh}\equiv -\bm{\tau}_{zh}\bm{\cdot}\nabla_z\widetilde{\bm{u}_h}$, $\Pi_{K,hz}\equiv -\bm{\tau}_{zh}\bm{\cdot}\bm{\nabla}_h\widetilde{{u}_z}$, $\Pi_{K,zz}\equiv -\bm{\tau}_{zz}{\nabla}_z\widetilde{{u}_z}$, $\Pi_{P,hh}\equiv -\bm{\Sigma}_h\bm{\cdot}\bm{\nabla}_h\widetilde{\phi}$, $\Pi_{P,zz}\equiv -{\Sigma}_z{\nabla}_z\widetilde{\phi}$. These $\Pi$ terms all correspond to energy transfer terms that describe the cascade of TKE and TPE to the sub-grid field. In the equations above we have dropped terms of order $\alpha^2$ and higher, and $c_1,c_2,c_3$ are used as short-hand for the coefficients of the $O(\alpha)$ terms (whose explicit forms can in principle be determined using the asymptotic expansions), and whose magnitudes are all $O(1)$.

The equations above can be used to determine the scaling of the filtered flow variables, which will be subsequently used to determine the behavior of the TKE and TPE dissipation rates since they are connected to the filtered flow dynamics through the energy cascades. For this, we note first that all of the terms in angled brackets on the rhs of the equations involve zero-order terms from the perturbation expansion, and are therefore independent of $\alpha$. Next, we must have $\langle \widetilde{p}^{[0]}\bm{\nabla}_h\bm{\cdot}\widetilde{\bm{u}_h}^{[0]}\rangle<0$ and hence $\langle \widetilde{p}^{[0]}\nabla_z\widetilde{{u}_z}^{[0]}\rangle>0$ if there are to be fluctuations in the vertical direction since the forcing only acts in the horizontal directions, $\langle \widetilde{\bm{u}_h}^{[0]}\bm{\cdot}\widetilde{\bm{F}}_h^{[0]}\rangle>0$ since this is the only source of energy in the system, and the energy flux terms will be positive since the TKE and TPE energy cascades are downscale in three-dimensional stratified turbulence \citep{lindborg06a}. We must also have $\langle\widetilde{\phi}^{[0]}\widetilde{{u}_z}^{[0]}\rangle>0$ since this is the only source of TPE in the system that can balance the TPE flux terms. The signs of $c_1,c_2,c_3$ are not known. While they can be formally evaluated as solutions to PDEs that can be constructed using the asymptotic series, these equations cannot be solved analytically. Therefore, in what follows we will simply treat these coefficients as being positive, but the implication is that while our asymptotic analysis can tell us how the sub-leading terms scale, it cannot say whether the sub-leading terms involving $\alpha$ make a positive or negative contribution.

Based on the discussion above regarding the signs of the terms in the equations we obtain the scaling relations 
\begin{align}
\gamma &\sim O\Big(2+\zeta(1+\zeta\delta^2) -\alpha\Big),\\
\beta &\sim O\Big(1+\zeta -\alpha\Big),\\
\delta &\sim O\Big(\zeta^{-1/2}(1-\alpha)^{-1/2}\Big).
\end{align}
The scaling of $\zeta$ is not yet established, and while this can be established using the scaled continuity equation, care is required. This is because $\bm{\nabla}_h\bm{\cdot}\widetilde{\bm{u}_h}$ involves contributions from two horizontal gradients, each of whose contribution is $O(1)$ under the scaling being used. However, the sign of these two contributions may be opposite leading to $\|\bm{\nabla}_h\bm{\cdot}\widetilde{\bm{u}_h}\|\ll 1$. As a consequence, $\zeta$ need not be $O(1)$ and could in fact be small. Nevertheless, we do have the constraint that under the scaling $\|\bm{\nabla}_h\bm{\cdot}\widetilde{\bm{u}_h}\|\leq O(1)$ from which it follows from the continuity equation (with the asymptotic expansion in $\alpha$ applied) that $\zeta\leq O(1-\alpha)$. In the limit $\alpha\to 0$, if $0<\zeta\ll 1$ then the result above for $\delta$ would yield $\delta \gg 1$. This would imply strong anisotropy of the velocity and scalar gradients in the neutrally buoyant limit which is contrary to expectation, suggesting that instead the upper bound $\zeta\sim O(1-\alpha)$ should be used in the weakly stratified regime. In this case, the scaling results become
\begin{align}
\gamma &\sim O\Big(4-2\alpha\Big),\label{gamma_result}\\
\beta &\sim O\Big(2 -2\alpha\Big),\label{beta_result}\\
\delta &\sim O\Big(1+\alpha\Big),\label{delta_result}\\
\zeta&\sim O(1-\alpha),\label{zeta_result}
\end{align}
where we have dropped the contribution from all terms smaller than $O(\alpha)$, which will also be done in the analysis that follows. 

Re-arranging the definitions of $\zeta,\gamma,\beta,\delta$ we obtain 
\begin{align}
\mathcal{U}_{h} &= \gamma^{-1/3}U_h(\ell_h/L_{h})^{1/3},\label{Uh_WS}\\
\mathcal{U}_{v} &=\delta \zeta \gamma^{-1/3}U_h(\ell_h/L_{h})^{1/3},\label{Uv_WS}\\
\mathcal{Q} &=\beta^{-1/2}\gamma^{1/6} U_v^{1/2} Q^{1/2} F_h^{-1/2}(\ell_h/L_{h})^{1/3},\\
\mathcal{F}_h&=\gamma^{-1/3} F_h(\ell_h/L_{h})^{-2/3},\\
\mathcal{R}_h&= \gamma^{-1/3} R_h(\ell_h/L_{h})^{4/3},\\
\alpha &= \delta^{-1}\zeta^{-1}\gamma^{2/3} Q U_h^{-1} F_h^{-1}(\ell_h/L_{h})^{1/3},
\end{align}
where ${R}_h\equiv L_h {U}_{h}/\nu$, ${F}_h\equiv U_h/(L_h N)$. Substituting into these expressions the scaling results from \eqref{gamma_result} through \eqref{zeta_result} then leads to results that show how the relationships between $\mathcal{U}_{h}, \mathcal{U}_{v}, \mathcal{Q}$ and  $U_h, U_v, Q$ depend on $\alpha$. However, while $U_h$ and $L_h$ are considered input parameters (since they are determined by the imposed forcing), $U_v$ and $Q$ are emergent variables, and so the scaling analysis must relate these to $U_h$ and $L_h$ to be complete. These are determined from the results above for $\mathcal{U}_{v}, \mathcal{Q}$ by using the conditions that were discussed earlier, namely, for $\ell_h\sim O(L_h)$ we must have $\mathcal{U}_{v}\sim O(U_v)$ and $\mathcal{Q}\sim O(Q)$. Using these conditions in the results above for $\mathcal{U}_{v}, \mathcal{Q}$ then leads to
\begin{align}
U_v &\sim O\Big([\delta \zeta \gamma^{-1/3}]_L U_h\Big),\\
Q &\sim O\Big([\beta^{-1}\delta \zeta]_L F_h^{-1} U_h\Big),\label{Qlarge_WS}
\end{align}
where $[\cdot]_L$ denotes that the variables inside the brackets are evaluated at $\ell_h=L_h$. We have therefore now completely determined how the variables $\mathcal{U}_{h}, \mathcal{U}_{v}, \mathcal{Q}$ depend on $U_h, L_h$ and $\alpha$. These will now be used to determine the scaling of the TKE and TPE dissipation rates.

From the dimensional form of the small-scale TKE and TPE equations we have for a statistically stationary and homogeneous flow \citep{zhang22}
\begin{align}
0&=-\langle \Pi_K^*\rangle+2\nu\langle\widetilde{\|\bm{S}^*\|^2}-\|\widetilde{\bm{S}^*}\|^2\rangle +N\langle\widetilde{u_z^*\phi^*}-\widetilde{u_z^*}\widetilde{\phi^*}\rangle-\langle\widetilde{\bm{F}_h^*\bm{\cdot}\bm{u}_h^*}-\widetilde{\bm{F}_h^*}\bm{\cdot}\widetilde{\bm{u}_h^*} \rangle,\label{SS_TKE}\\
0&=-\langle \Pi_P^*\rangle+\kappa\langle\widetilde{\|\bm{\nabla}^*\phi^*\|^2}-\|\widetilde{\bm{\nabla}^*\phi^*}\|^2\rangle-N\langle\widetilde{u_z^*\phi^*}-\widetilde{u_z^*}\widetilde{\phi^*}\rangle,\label{SS_TPE}
\end{align}
where for convenience we have written the dissipation terms using the total strain-rate $\bm{S}^*$ which is based on the total velocity gradient tensor $\bm{\nabla}^*\bm{u}^*=\bm{\nabla}_h^*\bm{u}_h^*+\bm{e}_z\nabla_z^* u_z^*\bm{e}_z$ and also the total scalar gradient $\bm{\nabla}^*\phi^*=\bm{\nabla}_h^*\phi^*+\bm{e}_z\nabla_z^*\phi^*$. Moreover, $\Pi_K^*$ and $\Pi_P^*$ are the sub-grid TKE and TPE energy fluxes, respectively, which are defined as 
\begin{align}
\Pi_K^*&\equiv -\bm{\tau}_{hh}^*\bm{:}\bm{\nabla}_h^*\widetilde{\bm{u}_h}^*-\bm{\tau}_{zh}^*\bm{\cdot}{\nabla}_z^*\widetilde{\bm{u}_h}^*-\bm{\tau}_{zh}^{*}\bm{\cdot}\bm{\nabla}_h^*\widetilde{{u}_z}^{*}-{\tau}_{zz}^*\nabla_z^* \widetilde{u_z}^*,\\
\Pi_P^*&\equiv -\bm{\Sigma}_h^*\bm{\cdot\nabla}_h^*\widetilde{\phi}^*-{\Sigma}_z^*\nabla_z^*\widetilde{\phi}^*.
\end{align}
For a homogeneous turbulent flow $2\nu\langle\widetilde{\|\bm{S}^*\|^2}\rangle=\langle\epsilon^*\rangle$ and $\kappa\langle\widetilde{\|\bm{\nabla}^*\phi^*\|^2}\rangle=\langle\chi^*\rangle$. Therefore, the small-scale TKE and TPE dissipation rates in \eqref{SS_TKE} and \eqref{SS_TPE} can be written as $2\nu\langle\widetilde{\|\bm{S}^*\|^2}-\|\widetilde{\bm{S}^*}\|^2\rangle=\langle\epsilon^*\rangle-2\nu \langle\|\widetilde{\bm{S}^*}\|^2\rangle$ and $\kappa\langle\widetilde{\|\bm{\nabla}^*\phi^*\|^2}-\|\widetilde{\bm{\nabla}^*\phi^*}\|^2\rangle= \langle\chi^*\rangle-\kappa\langle\|\bm{\nabla}^*\widetilde{\phi}^*\|^2\rangle$. Using these results in the small-scale TKE and TPE equations above and re-arranging leads to expressions for $\langle\epsilon^*\rangle$ and $\langle\chi^*\rangle$ (for $\mathcal{R}_h\to\infty$)
\begin{align}
\langle\epsilon^*\rangle&=\langle \Pi_K^*\rangle -N\langle\widetilde{u_z^*\phi^*}-\widetilde{u_z^*}\widetilde{\phi^*}\rangle+\langle\widetilde{\bm{F}_h^*\bm{\cdot}\bm{u}_h^*}-\widetilde{\bm{F}_h^*}\bm{\cdot}\widetilde{\bm{u}_h^*} \rangle,\label{SS_TKEb}\\
\langle\chi^*\rangle&=\langle \Pi_P^*\rangle+N\langle\widetilde{u_z^*\phi^*}-\widetilde{u_z^*}\widetilde{\phi^*}\rangle.\label{SS_TPEb}
\end{align}
The scaling of the terms on the rhs of these expressions can all be determined in terms of the filtered flow variables, and this will then lead to scaling results for $\langle\epsilon^*\rangle$ and $\langle\chi^*\rangle$.

Based on the scaling of the velocity and density gradients, and the scaling of the sub-grid stress terms together with the results in \eqref{Uh_WS} - \eqref{Qlarge_WS} we obtain
%
\begin{align}
\langle\Pi_K^*\rangle &\sim O\Big((1 +\zeta+\delta^2\zeta^2+\delta^2\zeta^3)\gamma^{-1} L_h^{-1} U_h^3\Big),\label{Pi_K_scaling}\\
\langle \Pi_P^*\rangle &\sim O\Big(\beta^{-1}(1+\zeta)[\delta^2 \zeta^2 \beta^{-1}\gamma^{-1/3}]_L F_h^{-2} L_h^{-1} U_h^3  \Big).\label{Pi_P_scaling}
\end{align}
The quantity $N\langle\widetilde{u_z^*\phi^*}-\widetilde{u_z^*}\widetilde{\phi^*}\rangle$ is the small-scale buoyancy production term. Since the small-scale fluctuations are less than or equal in order to the fluctuations occurring at the smallest scales in the filtered flow we have the upper bound
\begin{align}
N\langle\widetilde{u_z^*\phi^*}-\widetilde{u_z^*}\widetilde{\phi^*}\rangle&\leq O\Big(N\mathcal{Q}\mathcal{U}_{v}\Big),\label{SS_buoyancy}     
\end{align}
and using the results in \eqref{Uh_WS} - \eqref{Qlarge_WS} we obtain
\begin{align}
N\langle\widetilde{u_z^*\phi^*}-\widetilde{u_z^*}\widetilde{\phi^*}\rangle&\leq O\Big(\beta^{-1/2} \delta\zeta \gamma^{-1/6}[\beta^{-1/2} \delta\zeta \gamma^{-1/6}]_L F_h^{-2}(\ell_h/L_h)^{2/3} L_h^{-1} U_h^3 \Big).\label{SS_buoyancy2}     
\end{align}
In the small-scale TKE equation, $\langle\widetilde{\bm{F}_h^*\bm{\cdot}\bm{u}_h^*}-\widetilde{\bm{F}_h^*}\bm{\cdot}\widetilde{\bm{u}_h^*} \rangle$ denotes the direct injection of small-scale TKE due to the forcing. As stated earlier, we assume that the forcing only acts on the filtered field and therefore $\langle\widetilde{\bm{F}_h^*\bm{\cdot}\bm{u}_h^*}-\widetilde{\bm{F}_h^*}\bm{\cdot}\widetilde{\bm{u}_h^*} \rangle= 0$, which is reasonable provided $\ell_h\ll L_h$. The condition $\ell_h\ll L_h$ is usually satisfied in DNS and is assumed throughout the analysis.

Inserting the results just obtained into \eqref{SS_TKEb} and \eqref{SS_TPEb}, and using \eqref{gamma_result} through \eqref{zeta_result} leads (after a lengthy exercise in algebra) to
\begin{align}
\langle\epsilon^*\rangle&\sim O\Big(( 1-F_h^{-2}(\ell_h/L_h)^{2/3})L_h^{-1}U_h^3\Big),\label{SS_TKE_asmp}\\
\langle\chi^*\rangle&\sim O\Big( F_h^{-2}L_h^{-1}U_h^3\Big),\label{SS_TPE_asmp}
\end{align}
%
where higher order terms have been dropped. Note that in these results we have set all numerical coefficients that are $O(1)$ equal to unity. From these we can construct the asymptotic prediction for $\Gamma\equiv \langle\chi^*\rangle/\langle\epsilon^*\rangle$  
\begin{align}
\Gamma\sim O\Big(F_h^{-2}( 1+F_h^{-2}(\ell_h/L_h)^{2/3})\Big).\label{Gamma_WS_1}
\end{align}
Since in view of \eqref{Uh_WS} we have $\mathcal{U}_h\propto \ell_h^{1/3}$ (i.e. the same as in isotropic turbulence), then using the same arguments as those in \S\ref{SSlimit} we can show that for $\ell_h\ll L_{h}$ we have $U_{h}\sim O(U_{h,0})$ and $L_{h}\sim (L_{h,0})$ and therefore the result for $\Gamma$ may be written as
\begin{align}
\Gamma\sim O\Big(Fr_h^{-2}( 1+Fr_h^{-2}(\ell_h/L_{h,0})^{2/3})\Big),\label{Gamma_WS_2}
\end{align}
where $Fr_h\equiv U_{h,0}/(L_{h,0}N)$ and (for future reference) $Re_h\equiv U_{h,0}L_{h,0}/\nu$ are the Froude and Reynolds numbers based on the horizontal r.m.s velocity and horizontal integral lengthscale of the flow. 

Although $\langle\epsilon^*\rangle$, $\langle\chi^*\rangle$, and $\Gamma$ are physically independent of the filter length $\ell_h$, the asymptotic predictions for them depend on $\ell_h$. This is simply a reflection of the fact that the analysis is developed in terms of the $\ell_h$-dependent parameter $\alpha$. The value of $\ell_h$ used in the analysis is arbitrary other than that it must be such that $\alpha\ll1$, and $\ell_h/L_h$ must be small enough to justify the neglect of the forcing in the small-scale TKE equation. To leading order we have $\alpha\sim O(Fr_h^{-2}(\ell_h/L_{h,0})^{1/3})$, and so provided $Fr_h^2\gg1$, the condition $\alpha\ll1$ is satisfied at all scales in the flow. If we assume that we can justifiably neglect the effect of forcing in the small-scale TKE equation provided $\ell_h\leq O(L_{h,0}/10)$ then using the upper bound the result for $\Gamma$ becomes 
\begin{align}
\Gamma\sim O\Big(Fr_h^{-2}( 1+Fr_h^{-2})\Big).\label{Gamma_WS_3}
\end{align}
The leading order behavior described by \eqref{Gamma_WS_3} is the same as that obtained by \citet{maffioli16b} who derived the result using simple estimates. The sub-leading contribution $+Fr_h^{-4}$ comes from $O(\alpha)$ terms in the expansion and may in fact be $-Fr_h^{-4}$ due to the fact that the sign of the $O(\alpha)$ terms are not known, as discussed earlier. Therefore, the analysis predicts how the sub-leading term will scale with $Fr_h$, but not its sign.

Finally, note that the analysis yields $\mathcal{R}_h\sim O(Re_h(\ell_h/L_{h,0})^{4/3})$, and therefore for finite $\ell_h/L_{h,0}$, the limit $\mathcal{R}_h\to\infty$ which is assumed in the analysis is satisfied provided that $Re_h\to\infty$.

\subsection{Strongly stratified regime}

The strongly stratified regime corresponds to the regime where $\alpha\gg1$ at some scales in the flow. In the weakly stratified case, $\alpha\ll1$ holds at all scales provided that $F_h^2\gg1$ since $\alpha$ decreases with decreasing scale. In the strongly stratified case, $\alpha\gg1$ may only be satisfied at a sub-set of scales because $\alpha$ is expected to decrease with decreasing scale (which the analysis will show). Depending on how large $Fr_h$ and $Re_b$ are, there may be a sub-set of scales where $\alpha\ll 1$, corresponding to the inertial sub-range in a strongly stratified flow (the conditions for which will be explored in detail in \S\ref{L_0_est}).

In the limits $\alpha\to \infty$ and $\mathcal{R}_h\delta^2\to\infty$ (which also implies $\mathcal{R}_h\to\infty$ since we expect $\delta\leq O(1)$ for a stably stratified flow) the dominant balance for \eqref{uz_eq} yields $\zeta^{-1}\delta^{-2}\sim O(\alpha)$, and in this limit the dependence on $\delta$ disappears from the set of equations \eqref{uh_eq} - \eqref{phi_eq}. This shows that in this limit the filtered equations possess the same self-similarity properties that the unfiltered, Boussinesq-Euler equations were shown to possess in \citet{billant01}. In particular, when $\mathcal{R}_h\delta^2\to\infty$ and $\alpha\to \infty$, the dimensional version of equations \eqref{uh_eq} - \eqref{phi_eq} are invariant under the group of transformations defined by $N\to N/\xi$, $z^*\to\xi z^*$, $u_z^{*[0]}\to \xi u_z^{*[0]}$, where $\xi\in\mathbb{R}^+$ is a constant. Following the arguments of \citet{billant01}, this symmetry group implies $\ell_v\propto 1/N$, and using simple dimensional considerations $\ell_v\sim O(\mathcal{U}_{h}/N)$, which leads to $\delta\sim O(\mathcal{F}_h)$. This means that in the limits $\mathcal{R}_h\delta^2\to\infty$ and $\alpha\to \infty$, for any horizontal scale $\ell_h$ where the velocity scale is $\mathcal{U}_{h}$, vertical motion on the scale $\ell_v\sim O(\mathcal{U}_{h}/N)$ must emerge with velocity scale $\mathcal{U}_{v}\sim O(\mathcal{U}_{h}/\alpha \mathcal{F}_h)$ and scale-dependent vertical Froude number $\mathcal{F}r_v\equiv \mathcal{U}_{h}/(\ell_v N)\sim O(1)$ since this is the only way the equations can be balanced. This behavior will occur at all scales $\ell_h$ in the flow at which the limits $\mathcal{R}_h\delta^2\to\infty$ and $\alpha\to \infty$ can be taken. 


When $\alpha$ and $\mathcal{R}_h\delta^2$ are large but finite, the contribution of sub-leading terms in the vertical momentum equation breaks the self-similarity property just discussed. The behavior in this case can be analyzed using perturbation theory, with the zeroth-order solutions corresponding to the solutions to the self-similar form of the equations that exists for $\alpha\to \infty$ and $\mathcal{R}_h\delta^2\to\infty$. While we could expand in both small-parameters $\alpha^{-1}$ and $\mathcal{R}_h^{-1}\delta^{-2}$, our interest is only in understanding the sub-leading buoyancy corrections. Therefore, for the strongly stratified regime we expand variables in the small-parameter $\alpha^{-1}$, e.g. $\widetilde{{u}_z}=\widetilde{{u}_z}^{[0]}+\alpha^{-1}\widetilde{{u}_z}^{[1]}+O(\alpha^{-2})$ and similarly for the other variables, where the superscript $[0]$ denotes the zeroth-order solution, and at each order in the expansion $\mathcal{R}_h\delta^2\to\infty$ is assumed.

Equations for the average large-scale horizontal TKE $E_{K,h}\equiv \langle\|\widetilde{\bm{u}_h}\|^2\rangle/2$, large-scale vertical TKE $E_{K,v}\equiv\langle|\widetilde{{u}_z}|^2\rangle/2$ and large-scale TPE $E_P\equiv \langle\widetilde{\phi}^2\rangle/2$ can be derived from \eqref{uh_eq}, \eqref{uz_eq}, and \eqref{phi_eq} (see \citet{zhang22}). Re-arranging these to give equations for the energy transfer terms, inserting into the right hand sides of the equations the perturbation expansions in $\alpha^{-1}$, then for a statistically stationary, homogeneous flow we obtain (for $\mathcal{R}_h\delta^2\to\infty$)
\begin{align}
\begin{split}
\langle\Pi_{K,hh}\rangle+\zeta\langle\Pi_{K,zh}\rangle=&(1+\zeta^2\delta^2)\langle \widetilde{p}^{[0]}\bm{\nabla}_h\bm{\cdot}\widetilde{\bm{u}_h}^{[0]}\rangle +\gamma\langle \widetilde{\bm{u}_h}^{[0]}\bm{\cdot}\widetilde{\bm{F}}_h^{[0]}\rangle+d_1\alpha^{-1},
\end{split}
\\
\begin{split}
\langle\Pi_{K,hz}\rangle+\zeta\langle\Pi_{K,zz}\rangle=&\Big(\frac{1}{\zeta\delta^2}+\zeta\Big)\langle \widetilde{p}^{[0]}\nabla_z\widetilde{{u}_z}^{[0]}\rangle -\alpha\lambda\langle\widetilde{\phi}^{[0]} \widetilde{{u}_z}^{[0]}\rangle+d_2\alpha^{-1},\label{TKE_vert_SS}
\end{split}
\\
\begin{split}
\langle\Pi_{P,hz}\rangle+\zeta\langle\Pi_{P,zz}\rangle=&\beta \langle\widetilde{\phi}^{[0]}\widetilde{{u}_z}^{[0]}\rangle+d_3\alpha^{-1},
\end{split}
\end{align}
where $\lambda\equiv U_v/\mathcal{U}_v$, and $\lambda\geq O(1)$ since the velocities do not increase with decreasing scale. In the equations above we have dropped terms of order $\alpha^{-2}$ and higher, and $d_1,d_2,d_3$ are used as short-hand for the coefficients of the $O(\alpha^{-1})$ terms (whose explicit forms can in principle be determined using the asymptotic expansions), and whose magnitudes are all $O(1)$.

Using the same arguments as were used in the weakly stratified case for the signs of the terms in these equations we obtain the scaling relations 
\begin{align}
\gamma &\sim O\Big(2+\zeta(1+\zeta\delta^2) -\alpha^{-1}\Big),\\
\beta &\sim O\Big(1+\zeta -\alpha^{-1}\Big),\\
\delta &\sim O\Big(\zeta^{-1/2}\alpha^{-1/2}\lambda^{-1/2}\Big),\label{delta_SSa}
\end{align}
where higher-order terms have been dropped. As discussed earlier, for $\alpha\to \infty$ and $\mathcal{R}_h\delta^2\to\infty$, the filtered equations possess the same self-similar behavior as discussed in \citet{billant01}, and the scaling of the filtered equations should recover that of the unfiltered equations analyzed by \citet{billant01} when $\ell_h\sim O(L_h)$. This means that for $\ell_h\sim O(L_h)$ we should have $\zeta\sim O(1)$ to leading order, as found by \citet{billant01}. Since buoyancy becomes weaker with decreasing $\ell_h/L_h$, then $\zeta$ cannot decrease with decreasing $\ell_h/L_h$ since this would imply that the velocity gradients become increasingly anisotropic at smaller scales even though buoyancy is getting weaker. However, \eqref{cont_eq} enforces that $\zeta \leq O(1)$ to leading order, which then implies that we must have $\zeta\sim O(1)$ to leading order at all scales in the flow. Including the contribution in the continuity equation from the sub-leading term in the perturbation expansion we therefore have $\zeta\sim O(1-\alpha^{-1})$. 

Concerning $\lambda$, it was discussed earlier that the dominant balance for the vertical momentum equation gives the scaling $\zeta^{-1}\delta^{-2}\sim O(\alpha)$ for $\alpha\to\infty$. Since \eqref{TKE_vert_SS} is in fact derived from the vertical momentum equation, then consistency in the scaling of these two equations in the limit $\alpha\to\infty$ requires that $\lambda\sim O(1)$. This then also implies that $\mathcal{U}_v\sim O(U_v)$, i.e. $\mathcal{U}_v$ is independent of $\ell_h$ in the strongly stratified inertial range, unlike the weakly stratified inertial range where it is proportional to $\ell_h^{1/3}$. 

Using these results we obtain the simplified expressions

\begin{align}
\gamma &\sim O(3-\alpha^{-1}),\\
\beta &\sim O(2-2\alpha^{-1}),\\
\delta &\sim O(\alpha^{-1/2}),\label{delta_SSa}\\
\zeta &\sim O(1-\alpha^{-1}),
\end{align}
and using these in the results from equations \eqref{Uh_WS} through \eqref{Qlarge_WS} also leads to the leading order results
\begin{align}
\alpha &\sim O\Big(F_h^{-2}(\ell_h/L_h)^{2/3}\Big),\\
\mathcal{R}_h\delta^2 &\sim O\Big(R_h F_h^{2}(\ell_h/L_h)^{2/3}\Big).
\end{align}
Note that this implies that for $\alpha\gg1$, the scaling of the equations yields $\delta \sim O(F_h(\ell_h/L_h)^{-1/3})\sim O(\mathcal{F}_h(\ell_h/L_h)^{1/3})$. This shows that the result derived earlier $\delta \sim O(\mathcal{F}_h)$ which was based on the self-similarity of the equations in the limit $\alpha\to\infty$ together with simple dimensional analysis (as was done in \citet{billant01}) is missing the non-dimensional factor $(\ell_h/L_h)^{1/3}$ which is captured by a scaling analysis of the equations for $\alpha\gg1$. This can be interpreted as suggesting that the emergent vertical lengthscale in the stratified inertial range is not $\ell_v\sim O(\mathcal{U}_{h}/N)$ but rather $\ell_v\sim O((\ell_h/L_h)^{1/3}\mathcal{U}_{h}/N)$, which recovers the result of \citet{billant01} that $L_v\sim O(U_h/N)$ when $\ell_h\sim O(L_h)$.

Now that we have determined the scaling of $\gamma,\beta,\delta,\zeta$ in the regime $\alpha\gg 1$, these results can be inserted into \eqref{Pi_K_scaling}, \eqref{Pi_P_scaling}, and \eqref{SS_buoyancy2}, and then these inserted into \eqref{SS_TKEb} and \eqref{SS_TPEb}, leading to results for $\langle\epsilon^*\rangle$ and $\langle\chi^*\rangle$.
Finally, these can be used to obtain the asymptotic result for $\Gamma$
\begin{align}
\Gamma &\sim O\Big(1+Fr_h^2\Big),\label{Gamma_SS}
\end{align}
where as in the weakly stratified result, we have set all $O(1)$ numerical values to unity, have used $L_h\sim O(L_{h,0})$ and $U_h\sim O(U_{h,0})$ in view of \eqref{Uh_WS}, have set $\ell_h\sim O(L_{h,0}/10)$, and higher order terms have been dropped (see the weakly stratified analysis for a discussion of these steps). Once again, the sub-leading contribution $+Fr_h^{2}$ comes from $O(\alpha^{-1})$ terms in the expansion and may in fact be $-Fr_h^{2}$ due to the fact that the sign of the $O(\alpha^{-1})$ terms are not known, as discussed earlier. Therefore, the analysis predicts how the sub-leading term will scale with $Fr_h$, but not its sign. Note also that the analysis yields $\mathcal{R}_h\delta^2 \sim O(Re_b(\ell_h/L_{h,0})^{2/3})$, and therefore for finite $\ell_h/L_{h,0}$, the limit $\mathcal{R}_h\delta^2\to\infty$ which is assumed in the analysis is satisfied provided that $Re_b\to\infty$.

As a final comment on the analysis in this section, we note that \citet{maffioli16b} assume in their arguments and in the interpretation of their DNS results that provided the Taylor Reynolds number $Re_\lambda$ for the flow is sufficiently high (they assume $>200$ based on \citet{donzis05}), the TKE and TPE dissipation rates will have reached their asymptotic values which are approximately independent of $Re_\lambda$. This is not correct, however, and our analysis based on the filtering approach reveals why. For a weakly stratified turbulent flow, the anomalous behavior of the TKE dissipation rate is only established if $Re_h=O(Re_\lambda^2/15)\gg 1$ and $\ell_h\ll L_{h,0}$, so that viscous effects in the large-scale TKE equation are negligible, and forcing effects in the small-scale TKE equation to be negligible. When these conditions are satisfied the leading order asymptotic behaviour of the small-scale TKE equation is $\langle \Pi_K\rangle\sim(\ell_h/\mathcal{U}_{h}^3)\langle\epsilon^*\rangle$. With this asymptotic behavior, $(\ell_h/\mathcal{U}_{h}^3)\langle\epsilon^*\rangle$ is determined by the approximately inviscid filtered dynamics controlling $\langle \Pi_K\rangle$ and hence $(\ell_h/\mathcal{U}_{h}^3)\langle\epsilon^*\rangle$ must be approximately independent of $Re_h$. In a strongly stratified flow, however, the preceding analysis shows that viscous effects are only negligible in the large-scale TKE equation and forcing effects are only negligible in the small-scale TKE equation when $\mathcal{R}_h\delta^2\sim O(Re_b (\ell_h/L_{h})^{2/3})\gg 1$ and $\ell_h\ll L_{h,0}$. Since $Re_b\equiv Fr_h^2 Re_h$, having $Re_h\gg 1$ (or equivalently $Re_\lambda \gg 1$) does not guarantee that these conditions will be satisfied when $Fr_h\ll1$. Hence for a strongly stratified flow, whether the TKE and (by extension of these arguments) TPE dissipation rates will exhibit anomalous behavior is not determined by the size of $Re_\lambda$ but by the size of $\mathcal{R}_h\delta^2$.

\section{Relevance of the Ozmidov scale and the conditions for an inertial sub-range in strongly stratified flows}\label{Loz}

The Ozmidov scale $L_O\equiv (N^{-3}\langle\epsilon^*\rangle)^{1/2}$ is argued to denote the scale at which buoyancy and inertial forces are of the same order \citep{lesieur90,lindborg08}, and therefore that it is at scales smaller than $L_O$ that an inertial sub-range can emerge if $Re_b$ is large enough. Our scaling analysis of the filtered Boussinesq-Navier-Stokes equations can shed light on the correct interpretation of $L_O$ as well as clarify the conditions necessary for an inertial sub-range to emerge in strongly stratified turbulent flows.

In \citet{lindborg08}, the definition $L_O\equiv (N^{-3}\langle\epsilon^*\rangle)^{1/2}$ is derived by estimating the scale at which the TKE and TPE are of the same order. Their estimate is that the TKE at scale $\ell_h$ is given by $O(\ell_h^{2/3}\langle\epsilon^*\rangle^{2/3})$, and that the TPE at scale $\ell_h$ is given by $O(\ell_h^2 N^2)$. The value of $\ell_h$ at which $\ell_h^{2/3}\langle\epsilon^*\rangle^{2/3}=\ell_h^2 N^2$ gives the Ozmidov scale, $L_O\equiv (N^{-3}\langle\epsilon^*\rangle)^{1/2}$. According to our analysis, the estimate they use for the TPE is incorrect for a strongly stratified flow. Our scaling analysis suggests that the TPE at scale $\ell_h$ is actually given by $(1/2)\mathcal{Q}^2\sim O(\ell_h^{2/3}\langle\epsilon^*\rangle^{2/3})$ for a strongly stratified flow, i.e. the TKE and TPE are of the same order. This was predicted by \citet{billant01} to hold at the large scales, but our analysis shows that it holds at all scales $\ell_h$ at which the condition $\alpha\gg1$ is satisfied, such that there is no single scale at which the TKE and TPE are of the same order. On the contrary, we find that they are of the same order at all scales in the range $Fr_h^3L_{h,0}\ll\ell_h\ll L_{h,0}$ (the lower limit corresponding to the scale at which $\alpha\sim O(1)$). 

The scaling results $\mathcal{U}_{h}\sim O(\ell_h^{2/3}\langle\epsilon^*\rangle^{2/3})$ and $\langle\epsilon^*\rangle\sim O(U_{h,0}^3/L_{h,0})$ apply to leading order at all scales in the range $\eta_h\ll\ell_h\ll L_{h,0}$, and based on this we obtain $\mathcal{F}_h\sim O(Fr_h(L_{h,0}/\ell_h)^{2/3})\sim O(N^{-1}\ell_h^{-2/3}\langle\epsilon^*\rangle^{1/3})$. The value of $\ell_h$ at which $\mathcal{F}_h\sim O(1)$ is $\ell_h\sim O((N^{-3}\langle\epsilon^*\rangle)^{1/2})=O(L_O)$, and we also obtain $L_O\sim O(Fr_h^{3/2}L_{h,0})$ using $\langle\epsilon^*\rangle\sim O(U_{h,0}^3/L_{h,0})$. This shows that the correct interpretation of $L_O$ is not that it is the scale at which the TKE and TPE are of the same order (since this is satisfied for all scales in the range $Fr_h^3L_{h,0}\ll\ell_h\ll L_{h,0}$ and not at any particular scale), but that it is the scale at which $\mathcal{F}_h\sim O(1)$. That $L_O$ is the scale at which $\mathcal{F}_h\sim O(1)$ was also noted by \citet{lindborg08}.

Although $L_O$ is the scale at which $\mathcal{F}_h\sim O(1)$, this does not necessarily mean that it is the scale below which buoyancy forces are sub-leading. The small-scale buoyancy term is $N\langle\widetilde{u_z^*\phi^*}-\widetilde{u_z^*}\widetilde{\phi^*}\rangle$ and the scale below which this plays a sub-leading role is the scale below which this term is smaller than the order of the \emph{vertical} TKE flux terms, which is of order $O(\mathcal{U}_{\ell,v}^2\mathcal{U}_{\ell,h}/\ell_h)$. Using \eqref{SS_buoyancy2} and the scaling for the strongly stratified regime we find that to leading order
\begin{align}
\frac{\ell_h}{\mathcal{U}_{\ell,v}^2\mathcal{U}_{\ell,h}}N\langle\widetilde{u_z^*\phi^*}-\widetilde{u_z^*}\widetilde{\phi^*}\rangle\leq O(Fr_h^{-2}\ell_h/L_{h,0}).\label{L_0_est}
\end{align}
The scaling analysis for the strongly stratified regime shows that at all scales where viscous effects are sub-leading, the horizontal velocity and density fluctuations vary with $\ell_h$ as $\ell_h^{1/3}$. This decay rate with decreasing $\ell_h$ is fast enough to ensure that the sub-grid velocity and density fluctuations will be dominated by the largest scales in the sub-grid field (this is analogous to the argument that in isotropic turbulence, the fact that the velocity fluctuations decay as $\ell_h^{1/3}$ is fast enough to ensure that the TKE in the flow will be dominated by the largest scales, as was also shown in \S\ref{SSlimit}). Due to this, the upper bound in \eqref{SS_buoyancy2} and hence \eqref{L_0_est} should be used. Using this upper bound, then when $\ell_h=L_O\sim O(Fr_h^{3/2}L_{h,0})$ we find
\begin{align}
\frac{\ell_h}{\mathcal{U}_{\ell,v}^2\mathcal{U}_{\ell,h}}N\langle\widetilde{u_z^*\phi^*}-\widetilde{u_z^*}\widetilde{\phi^*}\rangle\sim O(Fr_h^{-1/2}).
\end{align}
Therefore, $L_O$ is not the scale at which buoyancy and the \emph{vertical} inertial terms are the same order, because at scale $\ell_h=L_O$ the buoyancy term is $O(Fr_h^{-1/2})$ larger than the vertical inertial terms, in the strongly stratified regime. The scale at which they are of the same order is the scale at which $Fr_h^{-2}\ell_h/L_{h,0}\sim O(1)$, that is $L_{O,v}\sim O(Fr_h^2 L_{h,0})\sim O(Fr_h^{1/2} L_O)$, which we may refer to as the vertical Ozmidov scale. 

For a strongly stratified flow, the importance of viscous effects at scale $L_{O,v}$ are determined by the size of
\begin{align}
\mathcal{R}_h\delta^2\Big\vert_{\ell_h=L_{O,v}}\sim O(Re_b Fr_h^{4/3}).\label{Reb_Lov}
\end{align}
Only if $Re_b Fr_h^{4/3}\gg1$ will there be an inertial sub-range at scales $\ell\ll L_{O,v}$ where both viscous and buoyancy forces are sub-leading compared with horizontal and vertical inertial forces. The result above shows that the condition for this is $Re_b Fr_h^{4/3}\gg1 $ not $Re_b\gg1$ as is usually thought (e.g. \citet{riley12}). The extra factor $Fr_h^{4/3}$ arises both because the relevant Reynolds number at scale $\ell_h$ is $\mathcal{R}_h\delta^2$ not $Re_b$ (and $\mathcal{R}_h\delta^2$ is a factor $(\ell_h/L_{h,0})^{2/3}$ smaller than $Re_b$), and also because the scale below which buoyancy is sub-leading compared to all inertial terms is $L_{O,v}$ not $L_O$.

The condition for there to be a range of scales where inertial forces are significant can also be expressed in another way. If there exists an inertial sub-range where the smallest scales are isotropic then at the Kolmogorov scale $\ell_h=\eta$ we have $\mathcal{F}_h=Gn^{1/2}$, where $Gn\equiv\langle\epsilon^*\rangle/(\nu N^2)$ is the ``activity parameter'' \citep[e.g.][]{debk19}. We use the symbol $Gn$ in recognition of Gibson's seminal work with this quantity and of Gargett's association of it with the dynamic range available in stratified flows for fully three-dimensional turbulence \citep{gibson80,gargett84}. If the smallest scales are in fact isotropic, then $\mathcal{F}_h=Gn^{1/2}$ will also indicate the importance of buoyancy on the vertical momentum equation at the Kolmogorov scale. In this case, in order for buoyancy forces to be small compared with inertial and viscous forces at the Kolmogorov scale (and thereby be consistent with the isotropic assumption) we require that $Gn^{1/2}$ be large enough to yield $\mathcal{F}_h\gg1$. Moreover, with the leading order scaling $\langle\epsilon^*\rangle\sim O(U_{h,0}^3/L_{h,0})$ that was derived earlier we obtain $Gn\sim O(Re_b)$. This shows that $Gn$ is related to both $\mathcal{F}_h$ at the Kolmogorov scale as well as to $Re_b$, and that the condition for small-scale isotropy is that $Gn$ is sufficiently high. The result in \eqref{Reb_Lov}, however, gives the more precise condition, namely that $Re_b\sim O(Gn)\gg Fr_h^{-4/3}$.

While the quantity $\langle\epsilon^*\rangle/(\nu N^2)$ is often also referred to as the buoyancy Reynolds number (e.g. as in \cite{maffioli16b}), and is predicted by the analysis to scale with $O(Re_b)$, it is not in general identically equal to $Re_b$. They are only equal when $\langle\epsilon^*\rangle=U_{h,0}^3/L_{h,0}$ is satisfied. Defining $A\equiv L_{h,0}\langle\epsilon^*\rangle/ U_{h,0}^3$, $A$ ranges between 0.4 and 1.81 in isotropic turbulence \citep{sreenivasan98}, and \citet{maffioli16} observed $A\gtrsim 0.3$ in their DNS of stratified turbulence. The average value for our DNS (details below) is $A \approx 0.25$. These values are sufficient to support the scaling relationship $\langle\epsilon^*\rangle\sim O(U_{h,0}^3/L_{h,0})$, but highlight that the actual values of $Gn$ and $Re_b$ will differ, with $Gn<Re_b$ usually.

\section{Direct Numerical Simulations}\label{DNS}

\subsection{Computation of Length Scales}

In \citet{maffioli16b}, the flow parameters are estimated and the predictions of the analysis are tested based on the assumption that $L_{h,0}\sim O(U_{h,0}^3/\langle\epsilon^*\rangle)$. In the context of our theoretical analysis, this relationship is a prediction from the theory, and to test the theory we should not assume a-priori that it is valid. Therefore, when computing $Fr_h, Re_h$ and $Re_b\equiv Fr_h^2 Re_h$ we use the actual integral lengthscale $L_{h,0}$ and horizontal r.m.s velocity $U_{h,0}$ computed from the DNS. The integral length scales are
computed as recommended in Appendix E of \citet{comte-bellot71}.

\subsection{Simulation Database}
The database used for this study consists of 29 simulations of forced,
homogeneous, stably stratified turbulence.  The simulations are motivated by
those of \citet{lindborg06a} and are of the same type as those previously
reported \citep{almalkie12a,debk15,portwood16}.  

\subsubsection{Numerical Method}
\label{sec:numerical}The numerical simulations for this study were computed using the same
methodology as those reported by \citet{almalkie12a}, \citet{debk15}, and
\citet{portwood16}, and the reader is referred to those papers for details. The only difference compared to these previous studies
is that in our DNS the viscous and diffusion terms are augmented
by fourth order hyperviscous and hyperdiffusive terms, with hyperviscosity $\nu_{hyp}$ and hyperdiffusivity $\kappa_{hyp}$. As shown below, these are negligible in
most of the simulations, but they are included in all simulations for consistency. In the simulations, $N$ is constant, $Pr=1$ and $\nu_{hyp}/\kappa_{hyp}=1$.

For the strongly stratified runs the horizontal forcing term $\bm{F}_h^*$ in the horizontal momentum equation converges the spectrum of
kinetic energy associated with horizontal motion, $E_h(\kappa_h, \kappa_z)$,
to a model spectrum for horizontal wave numbers $\kappa_h < \kappa_f$ and
vertical wave number $\kappa_z=0$, where $\kappa_f$ is eight times the
smallest non-zero wave number.  The forcing schema, denoted Rf in
\citet{rao11}, uses a spring-damper analogy to determine the input energy
needed as a function of $\kappa_h$ to quickly converge $E_h(\kappa_h, 0)$ to
the target spectrum, and then divides that input energy randomly among the
Fourier modes of the horizontal velocities with wave number $\kappa_h$ subject
to the constraint that continuity be satisfied.  A small amount of energy is
added stochastically to the horizontal velocities at $\kappa_h = 0$ and
$\kappa_z$ equal to 2, 3, and 4 times the smallest non-zero wave number.  The
model spectrum was determined by replicating Run 2 in \citet{lindborg06a}
using a stochastic forcing technique similar to that of \citet{alvelius99} and
denoted Qg in \citet{rao11}. 

The desired quasi-stationary simulation parameters $Fr_h, Gn$ were achieved by selecting
the mean density gradient and then adjusting $\nu =
\kappa$ to obtain the desired values.  The values of $\nu_{hyp}=\kappa_{hyp}$ were
chosen so as to maintain stability of the simulation having the highest
resolution requirement and using the largest numerical grid possible on the
computers available. Note that for the purposes of testing the current theoretical predictions it would have been desirable to control $Fr_h$ and $Re_b$ rather than $Fr_h$ and $Gn$, since it is the former pair of parameters that appear naturally in the scaling analysis. However, the existing DNS database had already been constructed based on controlling $Fr_h$ and $Gn$, and as we will show later, the DNS confirms that although the equality $Re_b=Gn$ does not hold, there is a clear scaling relationship between $Re_b$ and $Gn$ that is consistent with $Re_b\sim O(Gn)$.

\subsubsection{Spatial Resolution}
The resolution of the simulations is given by the number of grid points in the
horizontal ($N_x=N_y$) and vertical ($N_z$).  The large-scale spatial
resolution in the horizontal is the same for all the simulations because they
are forced to have a common target spectrum.  The large-scale spatial
resolution in the vertical scales with the Froude number \citep{billant01}.
To understand the dynamic range available for the inertial and dissipation
ranges, it is worthwhile to consider the history of DNS and the resolution
requirements for it.

The first three-dimensional DNS was performed in 1972
\citep{orszag72}, and the first of stratified turbulence in 1981
\citep{riley81}.  It was not until the late 1990's that simulations were
reported that are highly consistent with laboratory data for unstratified
turbulence and having sufficient dynamic range for an approximate inertial
range to exist \citep{wray97,debk98b,moser99}. Direct numerical simulations are traditionally defined as resolving ``all the
scales of motion.'' \citep{pope00}.  This might be possible for flows with low
Reynolds number, but it is usually impractical for resolving small-scale intermittent fluctuations in high Reynolds number flows. Therefore, a practical definition of DNS has
long been that the small length and time scales should be sufficiently resolved
so that the unresolved motions do not affect the dynamics of interest.
If one simply wants to ensure that $\langle\epsilon^*\rangle$ is well-resolved then the criteria $\kappa_{max}\eta > 1$
must be satisfied \citep{pope00}. If one is interested in resolving intermittent fluctuations then more recent studies conclude that $1.5 \le
\kappa_{max}\eta \le 3$ is the minimum resolution requirement for DNS, depending on the application
\citep[e.g.][]{zhou00c,yeung05,yakhot05,schumacher05,yeung06a,yeung06b,ishihara07,gulitski07b,gulitski07c,schumacher07b,schumacher07,watanabe07,donzis08,ishihara09,wan10,yeung18}.

In strongly stratified flows, there may be regions of relatively quiescent
flow (see \citet{portwood16} for images showing this in simulations comparable
to the current ones), so that $\eta$ based on the average dissipation rate is
larger than if it were calculated just for regions of strong turbulence.
Based on probability distribution functions of the local dissipation rates of
kinetic energy and density variance, \citet{debk15} concludes that $\kappa_{max}\eta >
3$ is required to resolve strongly stratified turbulence if internal
intermittency and dissipation-range dynamics are to be accurate.

While thumbrules based on $\kappa_{max}\eta$ are useful for estimating
resolution requirements, it is evident from the foregoing that a single
thumbrule may not be appropriate for flows spanning a wide range of Reynolds
and Froude numbers.  For this study, since we compute the hypervisous terms
even in highly resolved simulations, we can estimate the degree of small-scale
resolution directly by noting that the total dissipation rate is the sum of
the viscous and hyperviscous dissipation rates, $\epsilon^*_{T} = \epsilon^* +
\epsilon^*_{hyp}$, where $\epsilon^*_T$ is the total TKE dissipation rate for the flow, and $\epsilon^*_{hyp}$ is the contribution to the total TKE dissipation from the hyperviscous term.  When $\epsilon^*\approx \epsilon^*_T$, this indicates that the flow is well-resolved with respect to capturing the dynamics responsible for governing the TKE dissipation rate.

\subsubsection{Overview of Simulations}

\begin{table}
\begin{center}
\begin{tabular}{lrrrrrrrr}
\toprule
      & $Fr_h$ & $Gn$ & $Re_b$ & $Re_h$              & $N_x $ & $N_z $ & ${\cal L}_x $ & $\epsilon^* / \epsilon_T^*$ \\ \hline
 run1 &  0.039 &   14 &     58 & $ 3.9 \times 10^4 $ &  14784 &   1840 &             1 &                         1.0 \\
 run2 &  0.056 &   16 &     57 & $ 1.8 \times 10^4 $ &   6144 &    768 &             1 &                         1.0 \\
 run3 &  0.086 &   16 &     55 & $ 7.5 \times 10^3 $ &   2048 &    256 &             1 &                         1.0 \\
 run4 &  0.124 &   12 &     56 & $ 3.7 \times 10^3 $ &   2048 &    256 &             1 &                         1.0 \\
 run5 &  0.155 &   15 &     50 & $ 2.1 \times 10^3 $ &   2048 &    512 &             1 &                         1.0 \\
 run6 &  0.257 &   15 &     54 & $ 8.2 \times 10^2 $ &   1024 &    512 &             1 &                         1.0 \\
 run7 &  0.335 &   16 &     51 & $ 4.5 \times 10^2 $ &    512 &    256 &             1 &                         1.0 \\
 run8 &  0.449 &   15 &     27 & $ 1.4 \times 10^2 $ &    512 &    512 &             2 &                         1.0 \\
 run9 &  0.556 &   15 &     23 & $ 7.3 \times 10^1 $ &    256 &    256 &             2 &                         1.0 \\
run10 &  0.713 &   14 &     21 & $ 4.2 \times 10^1 $ &    128 &    128 &             2 &                         1.0 \\
\\run11 &  0.027 &   45 &    150 & $ 2.1 \times 10^5 $ &  18432 &   2304 &             1 &                         0.9 \\
run12 &  0.198 &   49 &    196 & $ 5.0 \times 10^3 $ &   4096 &   1024 &             1 &                         1.0 \\
run13 &  0.350 &   57 &    164 & $ 1.3 \times 10^3 $ &   4096 &   2048 &             1 &                         1.0 \\
run14 &  0.607 &   44 &    163 & $ 4.4 \times 10^2 $ &    512 &    512 &             2 &                         1.0 \\
run15 &  0.870 &   51 &    101 & $ 1.3 \times 10^2 $ &    256 &    256 &             2 &                         1.0 \\
run16 &  1.294 &   50 &     85 & $ 5.1 \times 10^1 $ &    128 &    128 &             2 &                         1.0 \\
\\run17 &  0.041 &  204 &   1521 & $ 9.1 \times 10^5 $ &  16384 &   2048 &             1 &                         0.2 \\
run18 &  0.105 &  194 &    543 & $ 4.9 \times 10^4 $ &  14784 &   1848 &             1 &                         1.0 \\
run19 &  0.209 &  202 &    868 & $ 2.0 \times 10^4 $ &  16384 &   4096 &             1 &                         1.0 \\
run20 &  0.322 &  207 &    729 & $ 7.0 \times 10^3 $ &   4096 &   2048 &             1 &                         1.0 \\
run21 &  0.648 &  202 &    689 & $ 1.6 \times 10^3 $ &    512 &    512 &             1 &                         1.0 \\
run22 &  1.438 &  197 &    521 & $ 2.5 \times 10^2 $ &    512 &    512 &             2 &                         1.0 \\
run23 &  2.578 &  206 &    443 & $ 6.7 \times 10^1 $ &    256 &    256 &             2 &                         1.0 \\
\\run24 &  0.026 & 1278 &   3200 & $ 4.7 \times 10^6 $ &  15840 &   1980 &             1 &                         0.1 \\
run25 &  0.210 & 1185 &   2862 & $ 6.5 \times 10^4 $ &  14784 &   3696 &             1 &                         0.7 \\
run26 &  0.419 & 1368 &   5263 & $ 3.0 \times 10^4 $ &  14784 &   7392 &             1 &                         1.0 \\
run27 &  0.564 &  929 &   2957 & $ 9.3 \times 10^3 $ &   3072 &   3072 &             1 &                         1.0 \\
run28 &  2.078 & 1044 &   5878 & $ 1.4 \times 10^3 $ &   1024 &   1024 &             1 &                         1.0 \\
run29 &  3.619 & 1066 &   5356 & $ 4.1 \times 10^2 $ &    512 &    512 &             1 &                         1.0 \\
\bottomrule
\end{tabular}
\end{center}
\caption{Parameters from the DNS simulations spanning strongly to weakly stratified flows. The rows are grouped according to their nominal $Gn$ values.  ${\cal L}_x$ is the size of the domain in the $x$-direction in units of $2\pi$. 
  \label{tbl:one}}
\end{table}
Parameters from the 29 simulations that comprise this study are tabulated in
table \ref{tbl:one}. The simulations fall into one of four sub-sets where the nominal value for $Gn$ is one of the four values $Gn \in \{14,\ 50,\ 200,\ 1000\}$. The numbers show that
most of the simulations are fully resolved with $\epsilon^*/\epsilon^*_T\approx 1$. For the cases where $\epsilon^*$ is appreciably smaller than $\epsilon^*_T$, the results can nevertheless still be of value provided 
$\epsilon^*/\epsilon^*_T$ is not too small. In particular, \citet{lalescu13} showed that turbulent motions in unstratified turbulence at scales $\lesssim 20\eta$ are slaved to the
chaotic motions of the larger scales.  Assuming Pope's model spectrum for
an integral-scale Reynolds number 10000, a grid resolution of $\approx 20\eta$
will resolve approximately $20\%$ of $\langle\epsilon^*\rangle$. Therefore, we estimate that cases where $\epsilon^*/\epsilon^*_T\gtrsim 0.2$ still provide meaningful information on the small-scale mixing, despite not being fully-resolved.

\FloatBarrier
\section{Results \& Discussion}\label{RandD}

\begin{figure}[h]
\includegraphics{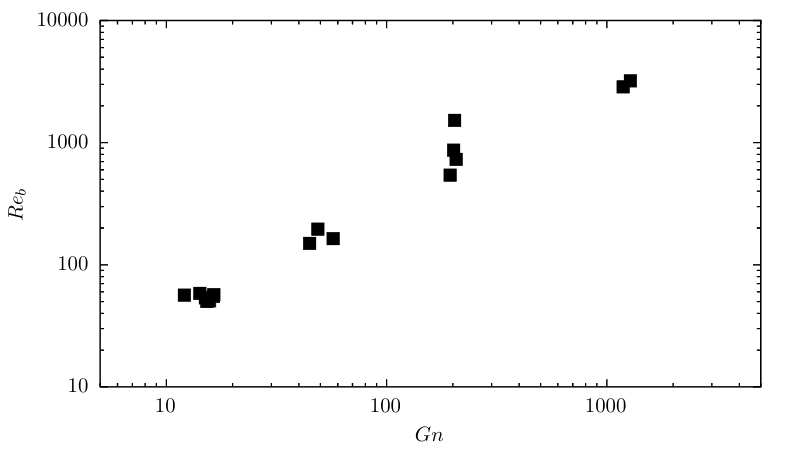}
\caption{Results to illustrate the relationship between $Re_b$ and $Gn$ in the DNS for the cases with $Fr_h < 0.4$. While $Gn$ is controlled in the DNS, $Re_b$ is not.
}
\label{Reb_Gn_plot}
\end{figure}

As discussed earlier, the scaling $\langle\epsilon^*\rangle\sim O(U_{h,0}^3/L_{h,0})$ is predicted to hold to leading order in weakly and strongly stratified flows and suggests that $Re_b\sim O(Gn)$. Since our analysis involves $Re_b$ (through its relation to $\mathcal{R}e\delta^2$), while the DNS results are based on controlling $Gn$, it is important to first check whether $Re_b\sim O(Gn)$ holds in the DNS to know to what extent results concerning the dependence of $\Gamma$ on $Gn$ might translate into results concerning the dependence of $\Gamma$ on $Re_b$. In figure \ref{Reb_Gn_plot} we plot $Re_b$ against $Gn$ for all of the DNS runs. The results show that while $Re_b$ is generally larger than $Gn$, and that for a given $Gn$ there may be a range of values of $Re_b$, there is a very clear relationship between the parameters in the sense that increasing $Gn$ corresponds to increasing $Re_b$. This then implies that it is reasonable to infer the dependence of $\Gamma$ on $Re_b$ from results showing the dependence of $\Gamma$ on $Gn$. 

We now turn to test the predictions concerning the asymptotic behavior of $\Gamma$ and its dependence on $Fr_h$ and $Gn$. As discussed in the introduction, one of the limitations of the DNS results in \citet{maffioli16b} for $\Gamma$ is that in their database $Fr_h$ and $Gn$ (what they call $Re_b$ is in fact $Gn$ not $Re_b\equiv Fr_h^2 Re_h$, and their $Fr_h$ differs from ours because theirs is based on estimating $L_{h,0}$ using $L_{h,0}\approx U_{h,0}^3/\langle\epsilon^*\rangle$) are in general varied simultaneously. As a result, it is impossible from their data to understand how $\Gamma$ depends on $Fr_h$ and $Gn$ distinctly. Our DNS are designed to avoid this issue by conducting runs where $Gn$ is approximately fixed while $Fr_h$ is varied, and doing this for different $Gn$.

%
%
%

\FloatBarrier

\begin{figure}
\includegraphics{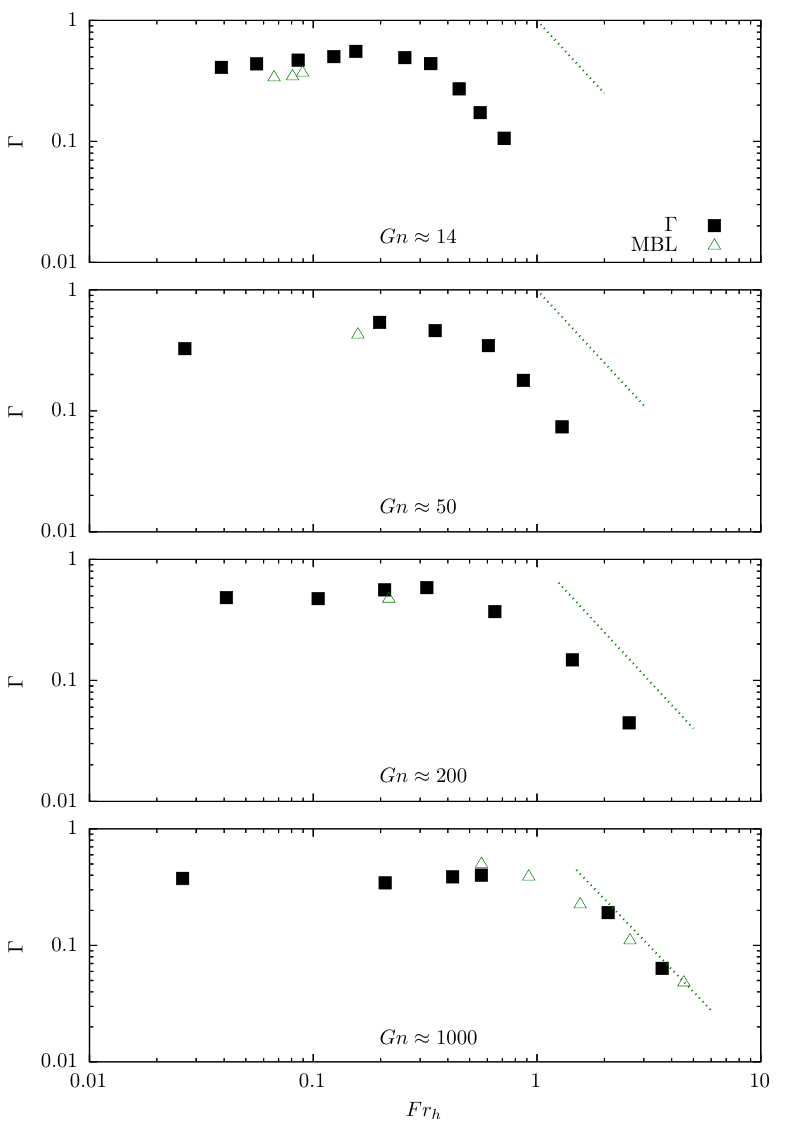}
\caption{DNS results for the mixing coefficient $\Gamma$ as a function of $Fr_h$. Each sub-panel corresponds to a different fixed value of $Gn$, with the black squares denoting data from our DNS, and the green triangles denoting data from the DNS of \citet{maffioli16b}. The green dashed line corresponds to $Fr_h^{-2}$ which is included to test the theoretical prediction that in the weakly stratified regime $\Gamma\sim O(Fr_h^{-2})$ to leading order.
\label{Gamma_Frh_plot}}
\end{figure}
%
%

\begin{figure}
\includegraphics{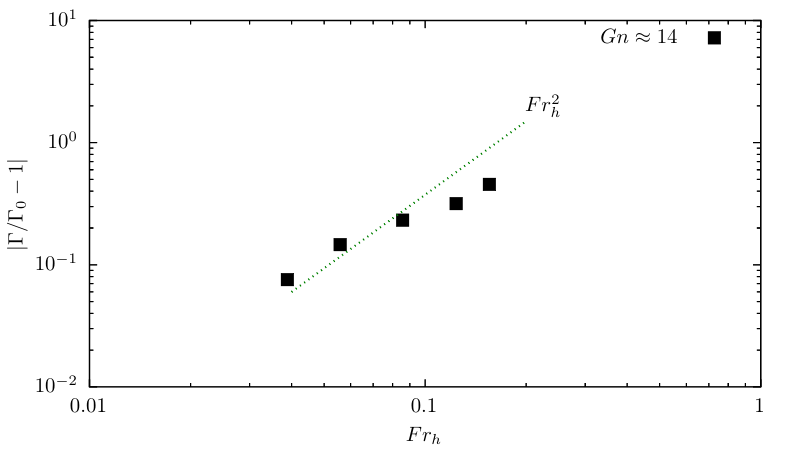}
\caption{DNS results to test the prediction of the theory that the sub-leading dependence of $\Gamma$ on $Fr_h$ scales as $\sim O(Fr_h^2)$, which implies $|\Gamma/\Gamma_0-1|\sim O(Fr_h^2)$, where $\Gamma_0\equiv \lim_{Fr_h\to 0}\Gamma$. In this plot, $\Gamma_0 \approx 0.44$ is estimated using a least squares fit to the data for $Gn\approx 14$. 
\label{Gamma_subleading}
}
\end{figure}

\begin{figure}
\includegraphics{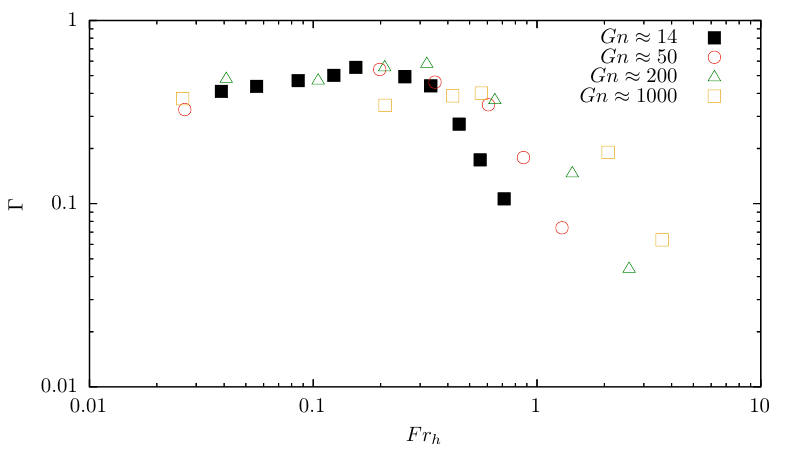}
\caption{DNS results for the mixing coefficient $\Gamma$ as a function of $Fr_h$, this time showing results for different $Gn$ on the same plot on order to see more clearly the effect of $Gn$.
\label{Gamma_Frh_plotb}
}
\end{figure}

In figure \ref{Gamma_Frh_plot} we plot the data for $\Gamma$ against $Fr_h$, where each sub-panel corresponds to a different value of $Gn$. In the plots, the black squares are data from our DNS, while the green triangles correspond to data from \citet{maffioli16b}, where we have transposed their data to be consistent with our definition of $Fr_h$ based on the computed integral length $L_{h,0}$. It can be seen that there is close agreement between our DNS results for $\Gamma$ and those of \citet{maffioli16b}, especially for $Gn\geq 50$.

The asymptotic prediction for the weakly stratified regime is $\Gamma\sim O(Fr_h^{-2}(1+Fr_h^{-2}))$, and in each sub-panel in figure \ref{Gamma_Frh_plot} the quantity $Fr_h^{-2}$ is shown as a dashed green line to test the leading order behavior $\Gamma\sim O(Fr_h^{-2})$. The results show that for the lower $Gn$ cases, the DNS values for $\Gamma$ are not equal in order of magnitude to $Fr_h^{-2}$, however, the results do show that $\Gamma\propto Fr_h^{-2}$. This is because no matter what the value of the flow Reynolds number, the buoyancy force will be proportional to $Fr_h^{-2}$ at the largest scales. The magnitude of the coefficient, however, will depend on $Gn$, and our theory effectively assumes $Gn\to \infty$, so the quantitative disagreement for lower values of $Gn$ is not surprising. However, the results show that for $Gn\approx 1000$, the leading order behavior $\Gamma\sim O(Fr_h^{-2})$ predicted by the asymptotic analysis is in excellent agreement with the DNS, with the DNS data for $\Gamma$ not merely following the functional dependence on $Fr_h$ predicted by the theory, but having values that are of the same order as predicted by the theory. 

For the opposite limit of strongly stratified turbulence, our results show behaviour that is consistent with the theoretical prediction $\Gamma\sim O(1+Fr_h^2)$ when $Fr_h\ll1$. Numerical evidence of this regime was already given in \citet{maffioli16b}, however, as mentioned previously, their data is somewhat hard to interpret because in it $Fr_h$ and $Gn$ vary simultaneously. Our results show that when $Gn$ is approximately fixed, $\Gamma\sim O(1)$ does indeed describe the correct leading-order asymptotic behavior. In this regime the data shows that there is a sub-leading dependence of $\Gamma$ on $Fr_h$, with $\Gamma$ generally slightly decreasing as $Fr_h$ decreases in the range $Fr_h\ll1$. The data is possibly consistent with the theoretical prediction that the sub-leading dependence on $Fr_h$ should scale as $\sim O(Fr_h^2)$, however, the data is not sufficiently converged in general to be able to clearly confirm this. The key reason for this is that the DNS for $Fr_h\ll1$ and $Gn\gg 1$ are very expensive to run, and running long enough simulations to fully converge the statistics in this regime is not currently feasible. The results for $Gn\approx 14$ are, however, sufficiently converged to approximately test the prediction from the theory for the sub-leading contribution to $\Gamma$. For this test we write the asymptotic prediction as $\Gamma\sim \Gamma_0+aFr_h^2$, where $\Gamma_0\sim O(1)$ and $a\sim O(1)$. We can then estimate $\Gamma_0\equiv \lim_{Fr_h\to 0}\Gamma$ using a least-squares fit to the data for $Gn\approx 14$ in the regime $Fr_h\ll1$, and the outcome is $\Gamma_0\approx 0.44$. Using this, in figure \ref{Gamma_subleading} we plot $|\Gamma/\Gamma_0-1|$, a quantity the theory predicts should scale as $\sim O(Fr_h^2)$. The results show that the DNS data for the three lowest $Fr_h$ values at $Gn\approx 14$ is approximately proportional to $Fr_h^2$. However, the data implies that the coefficient of proportionality is $a\gg1$, whereas the theory predicts that $a\sim O(1)$. This discrepancy is likely simply due to $Gn\approx 14$ being too small for the behavior to approximate the asymptotic behavior in the limit $Gn\to\infty$ that was assumed in the theory. This is reminiscent of the weakly stratified case where we saw that $\Gamma\propto Fr^{-2}$ holds at $Gn\approx 14$, but the coefficient of proportionality is not $O(1)$ as predicted by the theory.

In figure \ref{Gamma_Frh_plotb} we again plot the data for $\Gamma$ against $Fr_h$ but this time showing all the $Gn$ cases on one plot in order to see more clearly how the results depend on $Gn$ (We tried plotting the results as $\Gamma$ against $Gn$ with different sub-plots showing different $Fr_h$, however, because $Fr_h$ is not as controlled in the DNS as $Gn$, plotting the results in this way requires considering the results over sub-ranges of $Fr_h$ which then obscures the interpretation). The results show that for $Fr_h\geq O(1)$, $\Gamma$ is highly sensitive to $Gn$, the reasons for which have already been discussed. However, for $Fr_h\ll1$ the data for different $Gn$ approximately collapse, indicating that the leading order contribution to $\Gamma$ is weakly affected by $Gn$, even when $Gn$ is not large enough to be consistent with the behavior in the asymptotic limit $Gn\to\infty$.

Finally, as mentioned in the introduction, figure 4(b) of \citet{brethouwer07} shows DNS results that reveal that $\Gamma$ decreases significantly with decreasing $Re_b$ in the range $Re_b\leq O(1)$. This strong dependence of $\Gamma$ on $Re_b$ when $Re_b\leq O(1)$ is not merely of academic interest, but of practical importance for parameterizing $\Gamma$ since field observations in oceanic stratified flows show that $Re_b$ has a large range of values, spanning $O(10^{-2})\leq Re_b\leq O(10^5)$ (see figure 14 of \citet{jackson14}). The claim made in \citet{maffioli16b} that $\Gamma$ depends only on $Fr_h$ and not on the Reynolds number is therefore not in general correct but only true when $Re_b\gg1$ (which is possibly the only regime that \citet{maffioli16b} had in mind when making their argument). Stated precisely, in the weakly stratified regime, $\Gamma$ is independent of $Re_h$ in the limit $Re_h\to\infty$, in the strongly stratified regime, $\Gamma$ is independent of $Re_b\sim O(Gn)$ in the limit $Re_b\to\infty$, but in the strongly stratified regime with $Re_b\leq O(1)$, the results from figure 4(b) of \citet{brethouwer07} indicate that $\Gamma$ exhibits a leading order dependence on $Re_b$. It should be noted that this would occur even when $Re_h\gg1$ if $Fr_h$ is low enough such that $Re_h\leq O(Fr_h^{-2})$.


\section{Conclusions}\label{Conc}

This paper was motivated by the important study by \citet{maffioli16b} who considered the $Fr_h, Re_b$ dependence of the mixing coefficient $\Gamma$ in stratified turbulent flows. Using a simple scaling analysis they argued that in the weakly stratified flow regime $Fr_h\gg1$, $\Gamma\sim O(Fr_h^{-2})$ should hold. They conducted an extensive set of DNS of stratified turbulence, and the results for $Fr_h\gg1$ confirmed the scaling prediction $\Gamma\sim O(Fr_h^{-2})$. Their DNS results also indicated that in the strongly stratified regime $Fr_h\ll 1$, $\Gamma\sim O(1)$ holds, although their study did not provide a theoretical explanation for this. Their study also claimed that $\Gamma$ should in general depend on $Fr_h$ but should be independent of $Re_b$. Conclusive evidence for this was not given since in their DNS data set $Fr_h$ and $Re_b$ were varied simultaneously (except for a sub-set of results in the weakly stratified regime), so that the relative dependence of $\Gamma$ on $Fr_h$ and $Re_b$ could not be discerned. The study of \citet{maffioli16b} therefore left open two significant questions. First, to what extent does $\Gamma$ depend on $Fr_h$ as opposed to $Re_b$? Second, how can the result $\Gamma\sim O(1)$ observed in their DNS for $Fr_h\ll1$ be understood on theoretical grounds? 

To answer the first question, we used our DNS database of stratified turbulence where $Re_b$ is approximately fixed (actually it is the activity parameter $Gn$ that is fixed, but $Re_b$ and $Gn$ are proportional, as shown in \S\ref{RandD}) while $Fr_h$ is varied, for a wide range of values of $Re_b$. This allows us to clearly demonstrate the separate dependence of $\Gamma$ on $Fr_h$ and $Re_b$. The study of \citet{Garanaik19} sought to answer the second question and presented a simple scaling analysis that predicts $\Gamma\sim O(1)$ for $Fr_h\ll1$, consistent with the  DNS results of \citet{maffioli16b}. However, as discussed in the introduction, the scaling analysis of \citet{Garanaik19} seems problematic, and involves estimates for the TKE dissipation rate that are fundamentally inconsistent with well-established results for strongly stratified turbulence. In view of these issues, to answer the second question we developed a new asymptotic analysis of $\Gamma$ that predicts its dependence on $Fr_h$ in the limit $Re_b\to\infty$. One of the regimes of interest for the analysis is the strongly stratified turbulence regime where $Fr_h\ll1$ and $Re_b\to\infty$. The seminal study of \citet{billant01} explored the dynamics of stratified flows in the regime $Fr_h\ll1$ for inviscid fluids and discovered a new scaling regime that arises due to an emergent self-similarity of the flow in this regime. \citet{brethouwer07} extended the analysis to the case of viscous fluids and argued that when $Re_b\gg 1$ the behavior for $Fr_h\ll1$ reduces to the self-similar scaling regime identified by \citet{billant01}. However, we argued that this conclusion is problematic because the limit $Re_b\to\infty$ is singular. We therefore instead performed the asymptotic analysis on the filtered Boussinesq-Navier-Stokes equations in the strongly stratified turbulent regime, which allowed the singular limit $Re_b\to\infty$ to be handled correctly. This analysis reveals the precise sense in which the inviscid scaling analysis of \citet{billant01} applies to flows where viscous effects are important at the small-scales. Since the TKE and TPE dissipation rates are connected to the inter-scale TKE and TPE fluxes for a statistically stationary, homogeneous flow, the TKE and TPE dissipation rates could be obtained using expressions for the inter-scale TKE and TPE fluxes that are constructed from the filtered equations on which the asymptotic analysis was performed. This then allowed us to construct asymptotic predictions for $\Gamma$ in the limit $Re_b\to\infty$ for both the $Fr_h\ll1$ and $Fr_h\gg 1$ regimes.

For the weakly stratified regime $Fr_h\gg1$ we derived the prediction $\Gamma\sim O(Fr_h^{-2}(1+Fr_h^{-2})))$ which agrees to leading order with the result derived by \citet{maffioli16b}. For the strongly stratified regime our analysis predicts $\Gamma\sim O(1+Fr_h^{2})$ when $Fr_h\ll 1$. The leading order behavior $\Gamma\sim O(1)$ is consistent with the DNS results of \citet{maffioli16b}, and is also supported by our DNS. The sub-leading dependence of $\Gamma$ on $Fr_h$ observed in our DNS is consistent with the prediction of the theory that it should scale as $\sim O(Fr_h^2)$. However, the DNS data is not sufficiently converged to accurately test these predictions and this is due to the fact that DNS with $Fr_h\ll 1$ and $Re_b\gg 1$ are very expensive to run, and running them for long enough times for the sub-leading contributions to $\Gamma$ to be quantitatively testable is not currently feasible. We were able to show, however, that for a DNS with moderate $Re_b$, the sub-leading dependence of $\Gamma$ on $Fr_h$ is proportional to $Fr^2$ when $Fr_h\ll1$, in agreement with the theoretical prediction. The coefficient of proportionality is not $O(1)$, however, which is due to $Re_b$ being too low for the theory to truly apply.

For the strongly stratified regime $Fr_h\ll1$, DNS results from \citet{brethouwer07} show that $\Gamma$ exhibits a leading order dependence on $Re_b$ in the regime $Re_b\leq O(1)$. This, together with our asymptotic results suggests that the claim made by \citet{maffioli16b} that $\Gamma$ depends only on $Fr_h$ and not on the flow Reynolds number is not in general correct, but only true when $Re_b\gg1$ (which is possibly the only regime that \citet{maffioli16b} had in mind when making their argument). According to our analysis, in the weakly stratified regime, $\Gamma$ is independent of $Re_h$ in the limit $Re_h\to\infty$, in the strongly stratified regime, $\Gamma$ is independent of $Re_b$ in the limit $Re_b\to\infty$, but in the strongly stratified regime with $Re_b\leq O(1)$, the results from figure 4(b) of \citet{brethouwer07} indicate that $\Gamma$ exhibits a leading order dependence on $Re_b$. The latter behavior would occur even when $Re_h\gg1$ if $Fr_h$ is low enough such that $Re_h\leq O(Fr_h^{-2})$.

An important question for future work is to understand how the asymptotic predictions of our theory are modified when the Prandtl number is $Pr>1$, since our analysis assumed $Pr=O(1)$. Although this next step may seem simple, it is in fact a very complex question to address because we have recently shown that $Pr$ can have profound and surprising effects on the dissipation rates of TKE and TPE in stratified turbulent flows \citep{bragg2023understanding}. It is crucial to address, however, since in water flows, for example, salinity can lead to $Pr=O(1000)$. Another important question for future work is to understand how the asymptotic behavior of $\Gamma$ might differ when the flow is not driven by horizontal forcing but by a mean shear, something that has been explored in \citet{yi22}, as well as the more recent study \citet{yi23} that highlighted how the behavior of $\Gamma$ can depend on the type of forcing being used. An extension of our asymptotic analysis of the filtered Boussinesq-Navier-Stokes equations to cases with other kinds of forcing could provide insight into the parameter regimes of $Fr_h$ and $Re_b$ over which the asymptotic behavior of $\Gamma$ is sensitive to the nature of the forcing driving the flow.

\backsection[Acknowledgements]{This research used resources of the Oak Ridge Leadership Computing Facility at the Oak Ridge National Laboratory, which is supported by the Office of Science of the U.S. Department of Energy under Contract No. DE-AC05-00OR22725.  Additional resources were provided
through the U.S.\ Department of Defense High Performance Computing Modernization
Program by the Army Engineer Research and Development Center and the Army
Research Laboratory under Frontier Project FP-CFD-FY14-007.}

\backsection[Funding]{ADB was supported by National Science Foundation (NSF) CAREER award \# 2042346. SdeBK was supported by U.S. Office of Naval Research Grant number N00014-19-1-2152. 
}

\backsection[Declaration of interests]{The authors report no conflict of interest.}

\bibliographystyle{jfm}
\bibliography{bib}

\end{document}